\documentstyle[12pt]{article}
\def\beq{\begin{equation}}
\def\eeq{\end{equation}}
\def\bea{\begin{eqnarray}}
\def\eea{\end{eqnarray}}
\def\bq{\begin{quote}}
\def\eq{\end{quote}}

\def\beqa{\begin{eqnarray}}
\def\eeqa{\end{eqnarray}}
\def\be{\begin{equation}}
\def\ee{\end{equation}}
\def\beq{\begin{equation}}
\def\eeq{\end{equation}}

\def\bi{\begin{itemize}}
\def\ei{\end{itemize}}

\newcommand{\barre}[1]{%
        \setbox1=\hbox{$#1$} \dimen2=\ht1 \dimen3=\dp1 \dimen4=\wd1
        \setbox2=\hbox{\sl /}
        \dimen1=\wd1 \advance\dimen1 by -\wd2 \divide\dimen1 by 2
        \advance\dimen1 by \wd2 \advance\dimen1 by 0.4pt
        \setbox3=\hbox to \wd1{\hss \box1 \kern -\dimen1 \box2\hss}
        \ht3=\dimen2 \dp3=\dimen3 \wd3=\dimen4
        \box3
        }

\begin{document}
\pagestyle{empty}
\begin{flushright}   BONN-TH-99-17
\end{flushright}
\vskip 2cm
\begin{center}
{\huge Target-space Duality\\
in Heterotic and Type I 
Effective Lagrangians\\}
\vspace*{5mm} \vspace*{1cm} 
\end{center}
\vspace*{5mm} \noindent
\vskip 0.5cm
\centerline{\bf Zygmunt Lalak${}^{1,2}$
St\'ephane Lavignac${}^{1,3}$
and Hans Peter Nilles${}^{1}$}
\vskip 1cm
\centerline{\em ${}^{1}$Physikalisches Institut, Universit\"at Bonn}
\centerline{\em Nussallee 12, D-53115 Bonn, Germany}
\vskip 0.3cm
\centerline{\em ${}^{2}$Institute of Theoretical Physics}
\centerline{\em Warsaw University, Poland}
\vskip 0.3cm
\centerline{\em ${}^{3}$Service de Physique Th\'eorique, CEA-Saclay}
\centerline{\em F-91191 Gif-sur-Yvette C\'edex, France}
\vskip2cm

\centerline{\bf Abstract}
\vskip .3cm

We study the implications of target-space duality symmetries
for low-energy effective actions of various 
four-dimensional string theories. In the
heterotic case such symmetries can be incorporated in simple 
orbifold examples. At present a similar statement cannot be made
about the simplest type IIB orientifolds due to an obstruction at the
level of gravitational anomalies. This fact confirms previous 
doubts concerning a conjectured heterotic-type IIB orientifold 
duality and shows that target-space symmetries can be a powerful
tool in studying relations between various string theories at the
level of the effective low-energy action. Contraints on effective 
Lagrangians from these symmetries are discussed in detail. 
In particular, we consider ways of extending 
$T$-duality to include additional corrections to the K\"ahler potential 
in heterotic string models with $N=2$ subsectors.   


\newpage

\setcounter{page}{1} \pagestyle{plain}

\section{Introduction}

The idea of the string-driven 
unification of fundamental interactions 
has been given a new perspective with the discovery of the web of links,
usually referred to as dualities,  
possibly connecting different string theories. Evidence in favour 
of the existence of these links has been 
accumulated through the comparison of 
various compactifications of different ten-dimensional models and of 
eleven-dimensional supergravity. However,
for the application of these new developments to
phenomenologically relevant 
models in four dimensions,  progress 
has been made in two specific areas. First, the low-energy limit of the 
strongly coupled heterotic $E_8 \times E_8$ superstring and properties 
of its four-dimensional effective Lagrangian have been worked out in some 
detail \cite{witten}--\cite{ls_anom}.
Second, the novel class of four dimensional chiral type I models has 
been constructed through the compactification of the type IIB string
theory on six-dimensional orientifolds \cite{orientifolds}--\cite{lln}.

The type I--heterotic duality
in ten dimensions of Polchinski and Witten \cite{Polchinski}, 
which exchanges the strongly 
coupled sector of one theory with the weakly coupled sector of the other,
suggests then the existence of links between heterotic and type I 
models in lower dimensions. 
In addition, it has been realized that 
the volume of 
the compact six-dimensional space
appears as a new parameter relevant for duality symmetries.
This allows the possibility that heterotic--type I duality 
can link weakly coupled models on one side to  weakly coupled models 
on the other side in four dimensions. 
This has even lead to the concept of a 
generalized duality 
between heterotic orbifolds and type IIB orientifolds \cite{Sagnotti_Z_3},
relating models with different numbers of antisymmetric tensor fields.
And indeed, candidate dual models 
in the compactifications of the heterotic $SO(32)$ string have been found. 
However, since the underlying string theories look quite different, a
detailed comparison between type IIB orientifolds and heterotic models is
needed either to 
establish the conjectured duality firmly, or to diagnose points where it 
breaks down.

One obvious test of the 4d heterotic-type IIB orientifold 
duality is related 
to the 
presence of anomalous local $U(1)$ symmetries. In orientifold 
models several independent anomalous $U(1)$ factors may be present,
whereas 
in the 
explicitly 
known models on the
heterotic side there is always only a single anomalous $U(1)$.
Attempts to understand the physics of 
the presumably dual 
models with anomalous abelian factors have shed new light 
on the above duality conjecture, and in fact have lead us to identify
certain 
doubts on the validity of 4d heterotic-type IIB duality \cite{lln}.   
A nontrivial test of duality at the level of the 4d effective 
Lagrangians is associated with isometries of the 4d moduli spaces of 
prospective dual partners. The experience from the heterotic superstring 
models tells us that very often such isometries can be extended to
symmetries of the full classical effective Lagrangian. In the heterotic 
string  models the best known example of these symmetries are
target-space dualities, which are reflections of the underlying symmetry of 
string theory. As such, target-space dualities are expected to be good
quantum 
symmetries of the effective Lagrangian, which means that currents 
associated with them should be free of triangle gauge and gravitational 
anomalies. The requirement of exact quantum T-duality turns out to be a
powerful tool in studies of the four-dimensional string models. This 
symmetry restricts tree-level couplings in the effective action, 
and in addition allows us to determine the structure of one-loop
corrections 
to that action. 
This can be shown to be the case of threshold 
corrections to the gauge couplings in orbifold  models.
In the present paper we use target-space duality in this
spirit, to generalize the form of
one-loop corrections to the K\"ahler potential in the 
heterotic Lagrangian.

To decide whether the 4d heterotic--type IIB orientifold duality holds, 
one has to understand what exactly happens to target-space duality in the 
type IIB orientifold models. In addition, if target-space duality 
is there in the type IIB orientifold models, this might help 
to reconstruct the form of their  effective Lagrangians, which 
is crucial from the point of view of phenomenological applications.  
The problem of realization of target-space duality as a quantum 
symmetry in  effective Lagrangians which might describe type IIB
orientifold models in four dimensions is discussed in detail
in section 3 of the present paper.
We show that even in the simplest models with only D9-branes, it is
impossible to enforce cancellation of both gauge and
gravitational target-space duality anomalies by a Green-Schwarz mechanism.
Furthermore, even if one disregards the gravitational duality anomalies,
the structure of the recently computed threshold corrections
\cite{ABD} is not compatible with the Green-Schwarz cancellation of
gauge duality anomalies 
and seems to indicate that target-space duality does
not hold at the one-loop level in orientifold models.

However, there is more to say about the one-loop effective 
Lagrangian. 
In section 
4 we 
examine in detail specific heterotic models with respect to T-duality 
invariance. We explain that further corrections to the effective one-loop 
Lagrangian or to target-space duality 
transformations must arise in models with a plane fixed under 
some of the orbifold twists.
This necessity is due to those one-loop 
corrections 
to the K\"ahler potential which come together with the well known
holomorphic 
threshold corrections in models with $N=2$ subsectors.
The corrections that we discuss were not taken into account 
in the earlier analysis of modular anomalies.
In the version of the effective Lagrangian, where only holomorphic
thresholds 
are corrected to become covariant with respect to T-duality, the 
nonholomorphic corrections argued for in section 4 violate  
T-duality severely. They do it in a way that cannot be easily 
repaired without   
introducing additional kinetic mixing between 
$S$ and $T$ 
moduli. We propose such a modification, which generalizes the  
nonholomorphic corrections to the form which is invariant over the 
full range of the values of the $T$ modulus. The proper inclusion of the 
nonholomorphic corrections discussed in section 4 may modify 
somewhat phenomenological implications of the well-known heterotic models. 

\vskip 1cm


\section{Effective Lagrangians in heterotic models}

\subsection{Classical symmetries of moduli space}
Dimensional reduction of the ten-dimensional supergravity gives 
a nontrivial kinetic Lagrangian for the universal 
dilaton and geometric moduli superfields, 
which is invariant under the action of various 
symmetry transformations. 
Geometry of the moduli space is reflected by the  K\"ahler potential
\beq
K= -\log (S + \bar{S}) - \sum_{i=1}^{3} \log (T_i  + \bar{T}_i)
\label{e1} 
\eeq
There are two obvious symmetries\footnote{The symmetries which contain as 
a subset the invariances of the  kinetic Lagrangian of the moduli we shall 
refere to as sigma model symmetries.}
 of the kinetic 
part of the Lagrangian obtained by using this $K$ 
(for simplicity we put $T_1=T_2=T_3=T$ in the rest 
of this section): 
\beq \label{e2} 
T \rightarrow \frac{a T - i b}{i c T + d} \hskip 0.7cm
{\rm and} \hskip 0.7cm
S \rightarrow \frac{\tilde{a} S - i \tilde{b}}{i \tilde{c} S + \tilde{d}}.
\eeq
 The first symmetry, target-space duality, is 
believed to be, in its discrete form, the target-space version of 
a symmetry of the underlying 
string theory. The second is broken down at the level of the 
perturbative Lagrangian to an axionic shift through 
the coupling to the gauge fields. This target-space $S$-duality may be 
restored by nonperturbative effects, but this issue lies beyond the 
scope of the present paper.  

To make the target-space T-duality (\ref{e2}) the classical symmetry of the 
supergravity Lagrangian \cite{flat, lmn, flt, lidl}, 
one needs to transform the superpotential, W,
as well
\beq 
K \rightarrow K + 3 \log ( i c T + d ) + \, h.c. \; , \;  
W \rightarrow (i c T + d)^{-3} W  \label{eq:spp}
\eeq
If one switches off all the matter and nonuniversal moduli superfields, 
the suitable form of the  superpotential is 
$W (T) = \frac{const}{\eta^6 (T)}$ where $\eta^2$ is the squared Dedekind's 
 modular form of 
weight one. Further, it is possible to extend this clasical symmetry of the 
tree-level Lagrangian to 
include also matter and nonuniversal moduli fields. These fields transform 
as tensors, i.e. linearly, and their entries in the K\"ahler potential are
modular invariant  
on their own (see the Appendix). For instance, the K\"ahler potential for
untwisted matter is 
$K_A = A \bar{A} / (T + \bar{T})$, for twisted moduli C it turns out to be 
$K_C = C \bar{C} / (T + \bar{T})^3$, and for twisted matter 
$A_C$ $K_{A_C} = A_C \bar{A}_C / (T + \bar{T} )^2$ (like in the $Z_3$ orbifold
example).

\subsection{Cancellation of one-loop anomalies}

Target-space duality transformations involve sigma-model transformations,
and we have to check whether they are anomalous 
at the one-loop order \cite{Derendinger_sigma}, \cite{cardoso1},
\cite{cardoso2}. By the term sigma model we understand 
 the supersymmetric sigma model 
defined through the kinetic terms of the form $g_{i \bar{j}} (T, \bar{T}) 
\partial_\mu \phi^i \partial^\mu \bar{\phi}^{\bar{j}}$ where $g_{i \bar{j}} 
= \partial K / \partial \phi^i \partial \bar{\phi}^{\bar{j}}$ with the
K\"ahler potential{}\footnote{We suppress the dependence on the modulus $S$ as 
at tree level $S$ is inert under the $T$-duality transformations.} $K$. 
The form of the sigma model metric $g_{i \bar{j}}$ for indices corresponding 
to various fields was given at the end of the previous subsection. 
We can see, that upon $T$-duality transformations the form of $g_{i \bar{j}}$  
changes, and to compensate for that change one needs to `rotate' the  
$\phi$'s. These rotations, due to supersymmetry, act also on all fermions
present in the model, and transform them in the way of chiral rotations,
with charges related to the modular weights of the fields (see the 
Appendix). This chiral transformation results in an anomaly in the 
divergence of the current associated with $T$-duality. 
In addition, under $T$-duality the K\"ahler potential is not invariant,
but suffers a shift of the form $F(T) + \bar{F} (\bar{T})$ where $F$ is 
holomorphic. This shift is absorbed in the redefinition of the superpotential 
$W$, which in turn cancels against the transformation (\ref{eq:spp}) of $W$. 
Such K\"ahler transformation also results in the rotation of chiral fermions,
this time with the same charge for all chiral multiplets and just the opposite
one for gauginos. This rotation is also anomalous and the anomaly 
must be taken into account in addition to the sigma-model anomaly.
The anomalous  diagrams can 
be visualized as triangle diagrams with two gauge bosons or two gravitons 
and one composite connection which plays the role of the gauge field of 
$T$-duality. In fact, the part of the composite connection whose $T$-duality 
variation leads to nonvanishing variation of the diagrams is 
$V_\mu \sim \frac{\partial_\mu ( T - \bar{T} )}{ T + \bar{T}}$, 
since one typically assumes that all other non-inert fields fluctuate around 
vanishing vacuum values, hence 
the anomalous graphs are simply these with $\partial_\mu Im (T)$ at one vertex 
and gauge bosons or gravitons at remaining vertices. The sigma model 
anomalies can be computed in field theory limit of string theories,
and should be cancelled for the sake of the quantum exactness of the 
target-space dualities. The form of the one-loop terms needed in the effective 
Lagrangian to cancel anomalies can be worked out in field theory limit, 
but these terms can also be directly computed in string theory in various 
orbifold models, and there they have simply the status of one-loop 
corrections to the effective action, with their coefficients given from 
string theory. In the rest of this subsection we recapitulate 
briefly the status of these corrections in relevant classes of 
orbifold models. 

Let us start with the the simplest cases when all the orbifold 
planes are rotated by the orbifold twists (e.g. the $Z_3$ and $Z_7$
orbifolds). 
Then there are neither
 holomorphic 
threshold corrections, nor the associated  corrections
to the K\"ahler potential that will be described in
section 4, and all T-duality anomalies, gauge and 
gravitational, are cancelled through the universal four dimensional 
Green-Schwarz mechanism. 
More precisely, the 1-loop anomalies get cancelled by the shift of the 
dilaton superfield S: $S \rightarrow S - \frac{3 \delta_{GS}}{8 \pi^2} 
 \log (i c T + d) $. Then the dilaton K\"ahler potential is modified 
to $K = - \log ( S + \bar{S} - \frac{3 \delta_{GS}}{8 \pi^2} 
 \log (T + \bar{T}))$ which is T-duality invariant. 
The details of that cancellation are given in the Appendix. 
This four-dimensional Green-Schwarz mechanism \cite{GS}, \cite{dsw}, 
\cite{agnt} 
is sufficient 
to cancel all anomalies in 
models as e.g. the $Z_3$ and $Z_7$ orbifolds. 

In orbifold models with invariant planes, which contain 
N=2 subsectors, anomalies are no longer universal and  
one needs additional counterterms to cancel them 
completely \cite{Derendinger_sigma}. These counterterms 
 depend on moduli in a holomorphic way, and they are interpreted as 
holomorphic 
one-loop corrections to the tree-level gauge kinetic functions. 
In the most general case, 
using eq. (\ref{eq:L_nl}), (\ref{eq:threshold}) and (\ref{eq:b'_i_a}) 
from the Appendix, one can
rewrite the one-loop effective lagrangian for the gauge fields as\footnote{
For simplicity we suppress the dependence of the threshold corrections 
on the complex structure moduli $U$.}:
\begin{eqnarray}
  {\cal L}_{GK} & = & \frac{1}{4}\: \sum_a \int \! \mbox{d}^2 \theta\ W^a W^a\
  P_C \left\{\, \left[\, S + \bar S\  + \sigma + \bar \sigma\ 
-\ \frac{1}{8 \pi^2}\ \sum_i \, \delta^i_{GS}
  \ln (T_i + \bar T_i)\, \right] \right.  \nonumber \\
  & &  \left. -\ \frac{1}{8 \pi^2}\ \sum_i\,
  (b^{\prime i}_a - \delta^i_{GS})\, \ln \left[\, |\eta ( T_i)|^4
  (T_i + \bar T_i)\, \right]\, \right\}\ +\ \mbox{h.c.}
\end{eqnarray}
where $\sigma (T)$ is the holomorphic part of the universal one-loop threshold
correction invariant 
under $SL(2,Z)$ $T$-duality transformations and approaching $-\frac{1}{4 \pi} 
T$ when $T \rightarrow \infty$.  
The two terms in the bracket are separately modular invariant. From this
expression one can extract the evolution of gauge couplings, upon adding the
field-independent terms due to loops of massless charged states:
\begin{eqnarray}
  \frac{1}{g^2_a}\: \left|_{\mbox{1-loop}} \right. & = & \mbox{Re} S\
  + \mbox{Re} \sigma\ -\ \frac{1}{16 \pi^2}\ \sum_i \, \delta^i_{GS} \ln (T_i + \bar T_i)\
  -\ \frac{b_a}{16 \pi^2}\: \ln \frac{\mu^2}{M^2_H}  \nonumber \\
  & &  -\ \frac{1}{16 \pi^2}\ \sum_i\, b^{N=2}_{a, i}\, \ln \left[\,
  |\eta ( T_i)|^4 (T_i + \bar T_i)\, \right]
\end{eqnarray}
where $M_H$ is the (heterotic) string scale. This expression reflects already 
the knowledge of the explicit string computations of the nonuniversal 
threshold 
corrections to gauge couplings, as the 
coefficients $b^{N=2}_{a, i}$ are one-loop 
beta functions coming from $N=2$ subsectors associated with invariant planes.
The holomorphic part of the universal threshold correction -- 
$\sigma (T)$ -- remains to be determined for arbitrary $T$ 
via string calculations.
 
String results are naturally formulated in the linear 
multiplet formalism (see the Appendix), 
since the string degrees of freedom are antisymmetric tensor fields 
themselves and not the pseudoscalars dual to them. Later in this paper   
we shall try to find  relations between these string degrees 
of freedom 
and the chiral multiplets suitable for low-energy supergravity description
of the type IIB orientifold models.
Here we merely point out that the expression of the
one-loop string coupling $1/l$ in terms of the effective supergravity moduli
fields $\mbox{Re} S$ and $\mbox{Re} T_i$ (the one-loop chiral-linear duality
relation (\ref{eq:L_S_duality_loop})) is determined by the 
cancellation of sigma-model anomalies (see the Appendix).

We can now move on to the discussion of the $Z_3$ 
and $Z_7$
type IIB orientifold compactifications and their relation to heterotic 
compactifications.


\section{Target-space duality in $D=4$, $N=1$ type IIB $Z_N$ orientifolds}

Let us consider orientifolds \cite{orientifolds,gp} of type IIB string theory
compactified on a $T^6 / Z_N$ orbifold \cite{Sagnotti_Z_3}--\cite{Lykken}.
Consistency
of these models (absence of RR tadpoles that would spoil the UV finiteness of
the theory) requires the introduction of D-branes on which open strings
can end. If the orientifold projector is chosen to be the world-sheet parity
$\Omega$, one can have either only D9-branes or both D9-branes and D5-branes.
In addition, not all twists
leading to $N=1$ supersymmetry are allowed by tadpole conditions (for a
classification see ref. \cite{Ibanez_orientifolds}); the consistent $Z_N$
orientifolds that contain only 9-branes are the odd $N$ ($Z_3$ and $Z_7$)
orientifolds.

It has been noticed \cite{Ibanez_sigma} that the classical Lagrangian of
orientifolds with only D9-branes is invariant under
$SL(2,R)_{T_i}$ transformations. Indeed, the effective Lagrangian describing
the dynamics of the open string and untwisted closed string sectors of these
models, which can be obtained by reduction and truncation from the $D=10$
type I supergravity action, has the same structure as the Lagrangian of the
untwisted sector of heterotic orbifolds:
\begin{eqnarray}
  f_9 & = & S\ ,  \hskip 2cm  W\ \sim\ C^9_1\: C^9_2\: C^9_3\ ,  \\
  K & = & -\ \ln (S + \bar S)\ -\, \sum_i\ \ln \left( T_i + \bar T_i
  + |C^9_i|^2 \right)\ ,
\end{eqnarray}
where $f_9$ denotes the gauge kinetic function for gauge fields of the 
D9-brane sector, and $C^9_i$ is a generic (99) matter field associated with the
$i^{\rm th}$ complex plane. The addition of the twisted closed string states
(which are gauge singlets) does not spoil this invariance. In models containing
$5_i$-branes (5-branes wrapping on the $i^{\rm th}$ compact plane),
$SL(2,R)_{T_i}$ is explicitly broken by the gauge kinetic
function of the ($5_i 5_i$) gauge bosons, $f_{5_i} = T_i$, but modular
invariance with respect to the $j^{\rm th}$ compact plane still holds as soon
as no $5_j$-branes are present \cite{Ibanez_sigma}.

One may then ask whether target-space modular invariance is a good quantum
symmetry of $D=4$, $N=1$ type IIB orientifolds. Although it is not clear
what would be the origin of this symmetry in the underlying string
theory\footnote{In heterotic
compactifications, discrete modular transformations of the $T_i$ moduli
correspond to a fundamental quantum string symmetry, $T$-duality.
On the contrary, there is no such connection between the target-space
modular invariance observed at the level of the effective Lagrangian of
orientifolds and $T$-duality, since the latter exchanges Dirichlet and
Neumann boundary conditions, and therefore does not leave a given D-brane
configuration invariant.}, it is expected to hold at the one-loop level if
heterotic - type IIB duality is valid, because then sigma-model anomalies are
compensated for in the dual heterotic orbifold models (strictly speaking, this
argument does not apply to orientifolds containing 5-branes, since these models
do not seem to have any perturbative heterotic dual). Thus investigating
the possibility of cancelling sigma-model anomalies in orientifolds amounts to
testing duality. Also, the question of whether target-space modular invariance
is a good quantum symmetry in orientifolds is interesting on its own, even if
this string duality does not hold, because it could give us information on
the effective Lagrangian of those models, which are not so well known.
The purpose of this section is to investigate this question at the level of
the effective field theory, using the information given by recent string
computations.

The authors of ref. \cite{Ibanez_sigma} have proposed a mechanism for the
cancellation of sigma-model anomalies that is reminiscent of the way
Abelian gauge anomalies are compensated for in orientifolds
\cite{Ibanez_anomalous}. The
gauge group of orientifold models often contains several anomalous Abelian
factors $U(1)_i$. Their anomalies are cancelled by a generalized
\cite{Sagnotti_generalized} Green-Schwarz mechanism involving $RR$
twisted antisymmetric tensors $B^k_{\mu \nu}$ with appropriate couplings
to the gauge fields. In a more familiar chiral superfield language, those
antisymmetric tensors are described by their pseudoscalar duals $a_k$, which
lie in the same chiral multiplets $M_k$ as the scalars corresponding to the
blowing-up modes of the orientifold,
$a_k = \mbox{Im} M_k \! \mid_{\theta = \bar \theta = 0}$. The basic
ingredients of the generalized Green-Schwarz mechanism are a coupling of the
$M_k$ to the gauge fields,
\begin{equation}
  f_a\ =\ f_p\ +\ \sum_{k=1}^{[\frac{N-1}{2}]}\, s_{ak}\, M_k\ ,
\label{eq:f_a_tree}
\end{equation}
with $f_p=S$ for gauge group coming from 9-branes, and a shift of the twisted
axions $a_k$ under a $U(1)_i$ gauge transformation:
\begin{equation}
  M_k\ \rightarrow\ M_k\, +\, i\, \delta^k_i \Lambda_i\ .
\label{eq:M_k_shift}
\end{equation}
In eq. (\ref{eq:f_a_tree}), the sum goes over independent twisted sectors, and
for a twist $\theta^k$ with no fixed plane, the $M_k$ fields are
defined by $M_k = \frac{1}{\sqrt{N_k}} \sum_{f=1}^{N_k} M^f_k$, where
$\{ M^f_k \}_{f=1 \ldots N_k}$ are the states from the $k^{\rm th}$ twisted
sector, each of them living at one of the $N_k$ points that are fixed under
$\theta^k$
(for a twist leaving some plane unrotated, not all twisted states fit into
linear multiplets, see ref. \cite{Klein}). It is interesting to note that
the shift (\ref{eq:M_k_shift}) is a one-loop effect in the low-energy
effective field theory, although its string origin is a tree-level
Green-Schwarz coupling at the level of the orientifold.
From eq. (\ref{eq:f_a_tree}) and (\ref{eq:M_k_shift})
one obtains that mixed $U(1)G_aG_a$ anomalies are cancelled if the
following conditions are fulfilled:
\begin{equation}
  C^i_a\ =\ 8\, \pi^2 \sum_k\: c^k_a\, \delta^k_i\ .
\end{equation}
This Green-Schwarz cancellation of $U(1)$-gauge anomalies has been confirmed
by a string computation of the couplings $s_{ak}$ and $\delta^k_i$ \cite{ABD}.
Since the absence of $U(1)$-gravitational anomalies is also required for the
model to be consistent, it is natural to assume \cite{Ibanez_sigma} that they
are compensated for by the same mechanism (this proposal has been confirmed
in the recent paper \cite{scr_ser}), i.e. that the twisted axions $a_k$
couple to the $R \widetilde R$ term:
\begin{equation}
  -\, \frac{1}{4}\: \left( a + t_k\, a_k \right)\, R \widetilde R\ .
\label{eq:a_k_R2}
\end{equation} 
$U(1)$-gravitational anomalies must then satisfy the conditions:
\begin{equation}
  \frac{\mbox{Tr} X_i}{12}\ =\ 8\, \pi^2 \sum_k\: t_k\, \delta^k_i\ .
\label{eq:U1_grav}
\end{equation}
Contrary to the $s_{ak}$, the couplings $t_k$ have not been computed yet;
however, they can be extracted from eq. (\ref{eq:U1_grav}), since the
$\delta^k_i$ are known from ref. \cite{ABD}.

The authors of ref. \cite{Ibanez_sigma} have proposed that sigma-model
anomalies associated with complex planes that are rotated by all twists
are cancelled by a similar Green-Schwarz mechanism.
The most general possibility is that both $S$ and the $M_k$ shift under
$SL(2,R)_{T_i}$ transformations: $S \rightarrow S
- \frac{1}{8 \pi^2}\, \delta^{i,S}_{GS} \ln (i c_i T_i + d_i)$, $M_k
\rightarrow M_k - \frac{1}{8 \pi^2}\, \delta^{i,k}_{GS} \ln (i c_i T_i + d_i)$.
The sigma-gauge anomalies must then satisfy:
\begin{equation}
  b^{\prime i}_a\ =\ \delta^{i,S}_{GS}\ +\ \sum_k\, s_{ak}\,
  \delta^{i,k}_{GS}\ .
\label{eq:sigma_gauge_GS}
\end{equation}
Since there are generically as many Green-Schwarz parameters
$\delta^{i,S}_{GS}$ and $\delta^{i,k}_{GS}$ as anomalies,
it is always possible to cancel the sigma-gauge anomalies
by means of a Green-Schwarz mechanism.
Given the couplings $s_{ak}$, which are known from ref. \cite{ABD}, and the
anomaly coefficients $b^{\prime i}_a$, which are computed from the massless
spectrum, eq. (\ref{eq:sigma_gauge_GS}) determines uniquely the Green-Schwarz
parameters that are needed in order to ensure anomaly cancellation.
One finds \cite{Ibanez_sigma}
$\delta^{i,S}_{GS} = 0$ for all $Z_N$ orientifolds of ref.
\cite{Ibanez_orientifolds}, i.e. the dilaton does not play any role in the
cancellation of sigma-gauge anomalies. Although this last feature, which is
reminiscent of the mechanism of Abelian gauge anomaly cancellation, 
may appear to be suggestive, it should be stressed that
it is not possible to check the conjecture of a Green-Schwarz mechanism on
the basis of an analysis of the mixed gauge anomalies, since there are as
many potential counterterms as anomalies.
This situation is to be contrasted with the case of heterotic
orbifolds with no fixed planes, in which three
parameters $\delta^i_{GS}$, $i=1,2,3$ must cancel all anomalies, thus implying
several consistency relations that can be checked in explicit models
(namely the anomalies $b^{\prime i}_a$ must be gauge-group independent).


\subsection{Sigma-gravitational anomalies}

However, one can test this conjecture by considering sigma-gravitational
anomalies. Indeed, any shift of the universal and/or twisted axions under
target-space modular transformations induces, through the couplings
(\ref{eq:a_k_R2}), a variation of the Lagrangian $\delta {\cal L} =
\frac{\theta_i}{768 \pi^2}\, \bar b^{\prime i}_{grav} R \widetilde R$
that cancels part of the triangle anomaly $b^{\prime i}_{grav}$.
Assuming that the sigma-gauge anomalies are compensated for by a Green-Schwarz
mechanism, the shifts of the axions $a$ and $a_k$ under $SL(2,R)_{T_i}$ are
determined in an unambiguous way by eq. (\ref{eq:sigma_gauge_GS}), therefore
the corresponding coefficient $\bar b^{\prime i}_{grav}$ read,
given the fact that $\delta^{i,S}_{GS} = 0$ in all cases:
\begin{equation}
  \bar b^{\prime i}_{grav}\ =\ 24\, \sum_k\, t_k\, \delta^{i,k}_{GS}
\end{equation}
If a Green-Schwarz mechanism ensures the validity of target-space
duality at the one-loop level, then one
should have $b^{\prime i}_{grav} = \bar b^{\prime i}_{grav}$. It is
straightforward to check this relation in explicit models. For definiteness
we restrict ourselves to the odd order $Z_N$ orientifolds, which contain only
9-branes and do not have any fixed plane.
The values of the triangle anomaly
$b^{\prime i}_{grav}$ and of the Green-Schwarz contribution
$\bar b^{\prime i}_{grav}$ are displayed in Table 1, where 
$b^{\prime i}_M$ denotes the contribution of the $M_k$ fields to
$b^{\prime i}_{closed}$; since they transform non-linearly under
$SL(2,R)_{T_i}$, they have zero modular weights and $b^{\prime i}_M$
is just the number of $M_k$ fields ($b^{\prime i}_M = 27$ and $21$  for
$Z_3$ and $Z_7$ respectively). Following ref. \cite{Ibanez_sigma}, we also
give, for later discussion, the separate contributions of the open
and closed string states to the triangle anomaly:
\begin{eqnarray}
  b^{\prime i}_{open} & = & -\ \dim G\ +\ \sum_{\alpha}\,
  (1 + 2 n^i_{\alpha})\ ,
  \\
  b^{\prime i}_{closed} & = & 21\ +\ 1\ +\ b^{\prime i}_{mod}\ ,
\end{eqnarray}
where $b^{\prime i}_{mod}$ denotes the contribution of the other modulinos
than the dilatino, and by definition
$b^{\prime i}_{grav} = b^{\prime i}_{open} + b^{\prime i}_{closed}$.

\vskip .8cm
\begin{center}
\begin{tabular}{|c|c|c|c|c|}
\hline
& & & & \\
$\hskip .3cm \mbox{model} \hskip .3cm$
&$\hskip .3cm  b^{\prime i}_{open} \hskip .3cm$
&$b^{\prime i}_{closed}$ &$b^{\prime i}_{grav}$
&$\bar b^{\prime i}_{grav} \equiv 24 \sum_k t_k \delta^{i,k}_{GS}$ \\ 
& & & & \\
\hline
\hline
& & & & \\
$Z_3$ &$-10$ &$19 + b^{\prime i}_M$ &$9 + b^{\prime i}_M$ &$-18$ \\
& & & & \\
\hline
& & & & \\
$Z_7$ &$-6$ &$21 + b^{\prime i}_M$ &$15 + b^{\prime i}_M$ &$-6$ \\
& & & & \\
\hline
\end{tabular}
\vskip .5cm
\centerline{Table 1}
\end{center}   
\vskip .3cm

From Table 1 it is obvious that $\bar b^{\prime i}_{grav} \neq
b^{\prime i}_{grav}$, i.e. the Green-Schwarz mechanism alone cannot
compensate for both sigma-gauge and sigma-gravitational anomalies. However,
there may exist some other effect that would cancel the remaining anomaly
$b^{\prime i}_{grav} - \bar b^{\prime i}_{grav}$. This is actually suggested
by the string diagramatics relevant for
sigma-gravitational anomalies in open string models, as explained in
ref. \cite{Ibanez_sigma}. Indeed, only diagrams with
an open string loop (the annulus and Moebius amplitudes) can contribute to
sigma-gauge anomalies (as well as $U(1)$ anomalies), but in the case of
sigma-gravitational anomalies, there are additional contributions coming
from diagrams with a closed string loop (the torus and Klein bottle
amplitudes). Such diagrams are not present for sigma-gauge anomalies, because
they would be of higher order in the string coupling. The statement that
sigma-gravitational anomalies are cancelled is equivalent to the statement
that all four diagram topologies should sum up to zero, and that the field
theory limit of these string diagrams should contain both the triangle
anomaly (which one can split into two separate contributions,
$b^{\prime i}_{open}$ and $b^{\prime i}_{closed}$) and the corresponding
counterterms. The open string diagrams are expected to provide the
field theory anomalous tree graphs that are characteristic of a Green-Schwarz
mechanism, since these diagrams, when considered in the closed string (tree)
channel, are suggestive of a factorized form with closed string states
propagating in the cylinder (from the cancellation of sigma-gauge anomalies,
we know that those states must be twisted RR antisymmetric tensors). The
interpretation of the closed string diagrams is less obvious. The authors
of ref. \cite{Ibanez_sigma} assume that factorization is possible and that
these diagrams generate a one-loop mixing between the dilaton and the $T$
moduli that make up the sigma-model connection. However, this proposal
appears to be incompatible with the above field theory analysis, since it
would imply a one-loop\footnote{Although it has been argued \cite{Ibanez_sigma}
on the basis of string diagramatics that a one-loop mixing between the
dilaton and the $T$ moduli should not affect sigma-gauge anomalies, because
it would give a higher order contribution, this is not the case in the
effective field theory, where such a mixing would yield terms proportional
to $F^a \widetilde F^a$ in the variation of the Lagrangian, together with the
$R \widetilde R$ piece.} shift of the universal axion under $SL(2,R)_{T_i}$
transformations, which has already been excluded on the basis of sigma-gauge
anomaly cancellation (the same conclusion would hold for an additional mixing
between the twisted $M_k$ fields and the $T$ moduli). This suggests that the
possible counterterms originating from closed string loops are not
Green-Schwarz terms. The only alternative possibility we can think of would be
a $T_i$-dependent correction to the CP-odd $R^2$ terms with the appropriate
behaviour under $SL(2,R)_{T_i}$, i.e.
\begin{equation}
  {\cal L}_{CT}\ =\ \frac{1}{32 \pi^2}\ \mbox{Im}\, \Delta (T)\,
  R \widetilde R\ ,  \hskip .7cm \Delta (T)\
  \stackrel{SL(2,R)_{T_i}}{\longrightarrow}\ \Delta (T)\
  +\ \frac{b^{\prime i}_{grav} - \bar b^{\prime i}_{grav}}{24}\
  \ln (i c_i T_i + d_i)\ .
\end{equation} 
However, such corrections are not expected in models which do not possess any
$N=2$ sectors. Moreover, they cannot appear at the perturbative level because
of the Peccei-Quinn symmetries associated to the $T_i$ axions. Still one
cannot exclude the possibility that they are generated nonperturbatively.

Before concluding this subsection, let us comment more precisely on the
concrete proposal of ref. \cite{Ibanez_sigma}. There it was assumed that
the contribution of the open string sector to sigma-gravitational anomalies,
$b^{\prime i}_{open}$, was exactly compensated for by the $SL(2,R)_{T_i}$
shift of the $M_k$ moduli, whereas $b^{\prime i}_{closed}$ was taken care of
by closed string loop diagrams (interpreted as a mixing between the dilaton
and the $T$ moduli, which as explained above seems to be strongly disfavoured
by the field theory analysis). If this picture were correct, one would find
$b^{\prime i}_{open} = \bar b^{\prime i}_{grav}$ in all explicit orientifold
models. However, Table 1 shows that it is not verified in the $Z_3$ case,
although it accidentally holds in the $Z_7$ case.

We are therefore led to the following conclusion: either
modular invariance is not a good quantum symmetry in orientifold models, which
casts a new doubt on the validity of heterotic - type IIB duality; or it is
a good quantum symmetry and the anomalies cannot be cancelled by a pure
Green-Schwarz mechanism, even in models that do not possess any fixed planes.
One would then require $T_i$-dependent, holomorphic corrections to the $R^2$
terms, which however cannot arise perturbatively.


\subsection{One-loop corrections to the gauge couplings}

As stressed above, the consideration of field theory anomalies is not enough
to decide whether target-space duality is a good quantum symmetry of type IIB
orientifolds, mainly because of the large number of possible counterterms.
Although the cancellation of sigma-gravitational anomalies already appears
to be problematic, further tests of this symmetry are needed. The one-loop
corrections to gauge couplings that have been recently computed in type IIB
orientifolds \cite{ABD} give us the opportunity to perform such a test.
Indeed, the string result should agree with the one-loop corrections
computed in the effective supergravity theory. While the former depends on
the linear multiplets of the theory, the latter is expressed in terms of
chiral multiplets only; a duality transformation relates the two formulations.
The relations between the linear multiplets and their dual chiral multiplets
are defined order by order in perturbation theory. As is well known from the
heterotic case\footnote{Green-Schwarz cancellation of sigma-model anomalies
in the presence of several linear multiplets $L$ and $L_k$ has been described
in detail in ref. \cite{Klein}.}, the presence of Green-Schwarz counterterms
modifies these relations at the one-loop order (see eq.
(\ref{eq:L_S_duality_loop}) in Appendix B).
In the following, we shall then try to relate the string and supergravity
expressions for the one-loop gauge couplings through an explicit linear-chiral
duality transformation. The duality relations obtained in this way should
tell us whether the couplings needed for the cancellation of
sigma-model anomalies are indeed generated in orientifold
models.

Let us first establish the tree-level duality relations. The linear multiplets
we have
to deal with are the universal linear multiplet $L \sim (l, B_{\mu \nu}, \chi)$
and a model-dependent number of twisted linear multiplets $L_k \sim (m_k,
B_{k \mu \nu}, \chi_k)$, which describe the dilaton and the blowing-up modes,
respectively, as well as their antisymmetric tensor partner. The tree-level
Lagrangian for $L$ and $L_k$ has to reproduce the tree-level
gauge couplings (here we consider only gauge groups coming from the
$9$-brane sector) \cite{ABD}
\begin{equation}
  \frac{1}{g^2_a}\ =\ \frac{1}{l}\ +\ \sum_k\, s_{ak}\, m_k\ .
\label{eq:g_I_9_ABD_tree}
\end{equation}
Assuming no kinetic mixing between the different linear multiplets (this is
indeed the case in the basis in which formulae (\ref{eq:g_I_9_ABD_tree}) is
written), one finds:
\begin{equation}
  {\cal L}\ =\ \int \! \mbox{d}^4 \theta\ \Phi\, (\widehat L, \widehat L_k)\ ,
  \hskip 1cm  \Phi\, (\widehat L, \widehat L_k)\ =\ \ln \widehat L\
  -\ \sum_k\, \widehat L^2_k\ ,
\label{eq:L_k_lagr}
\end{equation}
where $\widehat L = L - 2\, \Omega$ and
$\widehat L_k = L_k + \sum_a s_{ak} \Omega_a$ are the modified, gauge invariant
multiplets. Following the standard procedure (see Appendix B), one then
obtains the tree-level duality relations:
\begin{equation}
  \frac{1}{\widehat L}\ =\ \frac{S + \bar S}{2}\ ,  \hskip 1cm
  \widehat L_k\ =\ \frac{M_k + \bar M_k}{2}\ ,
\label{eq:L_M_k_duality_tree}
\end{equation}
as well as the K\"ahler potential and gauge kinetic function that define the
Lagrangian describing the dual chiral superfields $S$ and $M_k$:
\begin{equation}
  f_a\ =\ S\ +\ \sum_k\, s_{ak}\, M_k\ ,  \hskip 1cm
  K (S, \bar S, M_k, \bar M_k)\ =\ -\, \ln (S + \bar S)\ +\ \frac{1}{4}\:
  (M_k + \bar M_k)^2\ .
\label{eq:M_k_lagr}
\end{equation}
One recovers, as expected, the gauge kinetic function (\ref{eq:f_a_tree}). Note
that, since we started from eq. (\ref{eq:g_I_9_ABD_tree}), which is the first
order in a perturbative expansion around the orientifold point $m_k=0$,
eq. (\ref{eq:L_k_lagr}), (\ref{eq:L_M_k_duality_tree}) and the K\"ahler
potential in eq. (\ref{eq:M_k_lagr}) are valid at leading order in $M_k$
only.

At the one-loop level, the Lagrangian (\ref{eq:L_k_lagr}) receives corrections
which modify the duality relations (\ref{eq:L_M_k_duality_tree}) and the
Lagrangian in the chiral basis (\ref{eq:M_k_lagr}). The loop-corrected
Lagrangian in the linear basis (respectively the chiral basis) should
reproduce the one-loop gauge couplings computed in ref. \cite{ABD}
(respectively the one-loop gauge
couplings computed in the effective supergravity theory \cite{KL}).
For the sake of simplicity, we restrict our discussion to the case of odd $N$
$Z_N$ orientifolds. These models do not possess any $N=2$ sectors, so that the
only way to compensate for sigma-model anomalies is the Green-Schwarz
mechanism, as discussed in the previous subsection. The one-loop gauge
couplings obtained from the string computation of ref. \cite{ABD} read:
\begin{equation}
  \left. \frac{1}{g^2_a\, (\mu^2)}\ \right|_{\mbox{1-loop}}\ =\ \frac{1}{l}\
  +\ \sum_k\, s_{ak}\, m_k\ -\ \frac{b_a}{16 \pi^2}\:
  \ln \frac{\mu^2}{M^2_I}\ .
\label{eq:g_I_9_ABD}
\end{equation}
The generic expression for one-loop gauge couplings in effective supergravity
theories read \cite{KL}:
\begin{eqnarray}
  \frac{1}{g^2_a (\mu^2)}\: \left|_{\mbox{1-loop}} \right. & = & \mbox{Re} f_a
  \mid_{\mbox{1-loop}}\
  -\ \frac{b_a}{16 \pi^2}\: \ln \frac{\mu^2}{M^2_{pl}}\  \nonumber \\
  & + & \frac{1}{16 \pi^2}\: \left[\, c_a\, K\ +\ 2\, C_2(G_a)\, \ln g^{-2}_a\
  -\ 2\, \sum_\alpha\, T(R_\alpha)\, \ln \det Z_{R_\alpha}\, \right]\ ,
\label{eq:g_I_KL}
\end{eqnarray}
where $Z_{R_\alpha}$ is the effective wave function normalization matrix for
the representation $R_\alpha$, and $c_a = \sum_\alpha T(R_\alpha) - C_2(G_a)$.
In the right hand side of eq. (\ref{eq:g_I_KL}), the functions $K$,
$g^{-2}_a$ and $Z_{R_\alpha}$ are truncated at tree-level,
while $f_a \mid_{\mbox{1-loop}}$ includes the one-loop string
corrections. Since formula (\ref{eq:g_I_KL}) is valid in any supergravity
effective theory, independently of its string origin, we can apply it to
the orienfold models we are considering. Then $f_a$ is given by its tree-level
expression (\ref{eq:M_k_lagr}) (the string threshold corrections
vanish in the absence of $N=2$ sectors),
$K = K (S, \bar S; M_k, \bar M_k) - \sum_i \ln (T_i + \bar T_i)$ where
$K (S, \bar S; M_k, \bar M_k)$ is given by (\ref{eq:M_k_lagr}), and
$Z_{R_\alpha} = \Pi_i (T_i + \bar T_i)^{n^i_{R_\alpha}}$. Eq. (\ref{eq:g_I_KL})
then becomes:
\begin{equation}
  \frac{1}{g^2_a (\mu^2)}\ =\ \mbox{Re} S\ +\
  \sum_k\, s_{ak}\, \mbox{Re} M_k\ -\ \sum_i \frac{b^{\prime i}_a}{16 \pi^2}\:
  \ln (T_i + \bar T_i)\
  -\ \frac{b_a}{16 \pi^2}\: \ln \frac{\mbox{Re} S\, \mu^2}{M^2_{pl}}\ ,
\label{eq:g_I_KL_2}
\end{equation}
up to terms that are suppressed both by a one-loop factor and a dependence
on $\mbox{Re} M_k$, which we can neglect. Note that the $T$-dependent
non-harmonic corrections (third piece in the right hand side of eq.
(\ref{eq:g_I_KL_2})) are related, through supersymmetry, to the anomalous
triangle diagrams associated with the sigma-model connection.

Assuming that it is possible to perform a linear-chiral duality transformation
at the one-loop level in a consistent way, one can obtain the one-loop
linear-chiral duality relations for $\widehat L$ and $\widehat L_k$,
as was done in ref. \cite{ABD}, by comparing the string
expression (\ref{eq:g_I_9_ABD}) with the effective supergravity expression
(\ref{eq:g_I_KL_2}). This gives:
\begin{eqnarray}
  \frac{1}{l} & = & \mbox{Re} S\ ,  \label{eq:l_S_duality}\\
  m_k & = & \mbox{Re} M_k\ -\ \frac{1}{16 \pi^2}\ \sum_i\,
  \delta^{i,k}_{GS} \ln (T_i + \bar T_i)\ -\ \frac{b_k}{16 \pi^2}\:
  \ln \left( \frac{\mbox{Re} S}{\Pi_i\, \mbox{Re} T_i} \right)^{\! 1/2}\ ,
\label{eq:m_M_duality_1}
\end{eqnarray}
where the coefficients $\delta^{i,k}_{GS}$ are defined by eq.
(\ref{eq:sigma_gauge_GS}) (with $\delta^{i,S}_{GS} = 0$ in all models), and
the coefficients $b_k$ are defined by $b^{N=1}_a\ =\ \sum_k\, s_{ak}\, b_k$.
The second piece in the right hand side of the duality relation
(\ref{eq:m_M_duality_1}) then comes precisely with the coefficients needed
for the Green-Schwarz cancellation of the sigma-gauge anomalies; however,
the third piece transforms under $SL(2,R)_{T_i}$ in the same way as
Green-Schwarz counterterms and seems to spoil anomaly cancellation.
The presence of this troublesome term can be traced back to the difficulty of
relating the string expression (\ref{eq:g_I_9_ABD}) to the supergravity
expression (\ref{eq:g_I_KL_2}) through a linear-chiral duality transformation,
as we discuss below.

The expression for the one-loop gauge couplings in the linear
basis, eq. (\ref{eq:g_I_9_ABD}), does not contain any term proportional to
$\ln (T_i + \bar T_i)$. This means that
the one-loop Lagrangian for the linear multiplets must contains a
term\footnote{As stressed before in the context of Abelian gauge anomalies,
the couplings $\delta^{i,k}_{GS}$ arise at the one-loop level in the
effective field theory, although they are presumably present at tree level
in the orientifold sense.}
\begin{equation}
  \Delta {\cal L}_{GS}\ =\ -\, \frac{1}{8 \pi^2}\: \sum_{i,k}\,
  \delta^{i,k}_{GS}\, \widehat L_k \ln (T_i + \bar T_i)
\label{eq:GS_counterterms}
\end{equation}
whose contribution to the gauge couplings exactly cancels the non-harmonic
corrections coming from the supersymmetric partners of the anomaly diagrams,
$- \sum_i \frac{b^{\prime i}_a}{16 \pi^2}\, \ln (T_i + \bar T_i)$. Note that
this Green-Schwarz mechanism is not enough to ensure $SL(2,R)_{T_i}$
invariance of the gauge couplings, since the upper scale of logarithmic
running, the string scale $M_I$, is not invariant on its own. Indeed, when
expressed in units of the Planck mass, $M_I$ depends on the $T_i$ moduli:
$M^2_I = \frac{\lambda_I M^2_{Pl}}{2\, {\rm Re} S}
= (\frac{{\rm Re} S}{\Pi_i {\rm Re} T_i})^{1/2}
\frac{M^2_{Pl}}{2\, {\rm Re} S}$ (we have used the identities
$\mbox{Re} S = V_I M^6_I / \lambda_I$ and $\mbox{Re} T_i = V^i_I M^2_I /
\lambda_I$ \cite{Ibanez_orientifolds}, where $\lambda_I$ is the
ten-dimensional string coupling and $V^i_I$ is the volume of the $i^{\rm th}$
compact torus in the string metric).
Now the addition of the Green-Schwarz counterterms (\ref{eq:GS_counterterms})
to the Lagrangian (\ref{eq:L_k_lagr}) leads to the modified duality relations
\begin{eqnarray}
  \frac{1}{\widehat L} & = & \frac{S + \bar S}{2}\ ,  \\
  \widehat L_k & = & \frac{1}{2}\ \left[\, M_k + \bar M_k\ -\
  \frac{1}{8 \pi^2}\ \sum_i\, \delta^i_{GS} \ln (T_i + \bar T_i)\, \right]\ ,
\label{eq:m_M_duality_2}
\end{eqnarray}
which do not agree with eq. (\ref{eq:m_M_duality_1}) and therefore yield
an expression for the gauge couplings in the chiral basis that does not fit
the supergravity expression (\ref{eq:g_I_KL_2}).

One can try to solve this problem by adding to the Lagrangian of the linear
multiplets, beyond the terms (\ref{eq:GS_counterterms}), the piece that
reproduces the fitted duality relation (\ref{eq:m_M_duality_1}), i.e.
\begin{equation}
  \Delta {\cal L}_{2}\ =\ \frac{1}{16 \pi^2}\: \sum_{k}\,
  b_k \widehat L_k \ln \left[ \widehat L\, \prod_i (T_i + \bar T_i) \right]
\label{eq:fitted_counterterms}
\end{equation}
($\Delta {\cal L}_{2}$ also modifies the duality relation between $\widehat L$
and $S$, but the corrections can be neglected, because it is proportional both
to a one-loop factor and to $m_k$). This reproduces the supergravity
expression (\ref{eq:g_I_KL_2}), but looks somewhat ad hoc; in particular, this
amounts to shift the upper scale of running to the invariant scale
$M^2_{Pl} / \mbox{Re} S$ and to absorb the residual, non-invariant logarithmic
term into the effective one-loop Lagrangian.
Note that the non-invariant part of $\Delta {\cal L}_{2}$ has the same form
as the Green-Schwarz counterterms $\Delta {\cal L}_{GS}$; as a result, this 
new term modifies the shift of the $M_k$ fields under $SL(2,R)_{T_i}$ and
only a part of the sigma-model anomalies is cancelled. Indeed, the 
modified K\"ahler potential obtained from the duality transformation
is a function of the combinations of chiral superfields corresponding to the
linear multiplets $\widehat L_k$;
then requiring invariance of $K$ amounts to modify the shift of the $M_k$ to:
\begin{equation}
  M_k\ \rightarrow\ M_k\ -\ \frac{1}{8 \pi^2}\: \left( \delta^{i,k}_{GS}
  - \frac{b_k}{2} \right)\, \ln (i c_i T_i + d_i)\ .
\end{equation}
The uncancelled anomaly is $b^{\prime i}_a - \sum_k s_{ak} (\delta^{i,k}_{GS}
- \frac{b_k}{2}) = b_a / 2$. Alternatively, one may absorb the terms
(\ref{eq:fitted_counterterms}) into a redefinition of the twisted linear
multiplets at the one-loop level,
$\widetilde L_k = L_k - \frac{b_k}{32 \pi^2}\, \ln\, [\widehat L\,
\Pi_i (T_i + \bar T_i)]$. The one-loop duality relations are then given by eq.
(\ref{eq:m_M_duality_2}), and the Green-Schwarz couplings ensure exact
cancellation of sigma-gauge anomalies. However, target-space duality is
violated at the level of one-loop corrections to the K\"ahler potential,
due to the fact that the constraints $D^2 L_k = D^2 (\widetilde L_k
- \frac{b_k}{32 \pi^2}\, \ln\, [\widehat L\, \Pi_i (T_i + \bar T_i)]) = 0$
and $\bar D^2 L_k = 0$ are no longer invariant.

We conclude that the structure of the one-loop corrections to gauge couplings
in type IIB orientifolds does not seem to be compatible with the cancellation
of sigma-gauge anomalies by a pure Green-Schwarz mechanism, even in the
simplest models with no $N=2$ subsectors, contrary to what the mere analysis
of anomalies would suggest. This, together with the difficulty observed at the
level of mixed gravitational anomalies, suggests that target-space duality is
merely an accidental, tree-level symmetry of the effective supergravity
Lagrangian of orientifold models, and strengthen previous doubts \cite{lln} on
the validity of the heterotic - type IIB duality. Note that it is not
completely
clear how the agreement between the string and effective supergravity
expressions for the one-loop gauge couplings is obtained; in this respect, an
explicit string computation of corrections to the twisted moduli K\"ahler
potential would give very useful information.


\subsection{Threshold corrections and heterotic - type IIB duality}

We have seen that target-space duality can be used as a tool to test
heterotic - type IIB duality at the one-loop level.
For completeness, we would like to mention another nontrivial check based
on the comparison of the threshold corrections in candidate dual
orbifold/orientifold models, which also raises doubts on the validity of
this duality (the following discussion is taken from ref. \cite{ABD}).
Specifically, we shall consider the $Z_3$ models, in which
a perfect matching of the massless spectra and gauge groups of both
low-energy effective field theories is obtained after decoupling of the
anomalous $U(1)$.
The one-loop gauge couplings of the $Z_3$ orbifold/orientifold computed in
string theory read \cite{ABD}:
\begin{eqnarray}
  \frac{1}{g^2_a (\mu^2) \mid_H} & = & \frac{1}{l_H}\
  -\ \frac{b^H_a}{16 \pi^2}\: \ln \frac{\mu^2}{M^2_H}\ ,  \\
  \frac{1}{g^2_a (\mu^2) \mid_I} & = & \frac{1}{l_I}\ +\ s_a\, m\
  -\ \frac{b^I_a}{16 \pi^2}\: \ln \frac{\mu^2}{M^2_I}\ ,
\label{eq:g_Z_ABD}
\end{eqnarray}
where $M_H$ (respectively $M_I$) is the heterotic (respectively type I)
string scale, and $m$ is the symmetric combination of the
27 twisted scalars. At
tree level\footnote{Needless to say, there is a perfect matching of the gauge
couplings of both models at tree-level. Indeed, at the point of maximal gauge
symmetry, one has $m=0$ (corresponding to $\xi^2_I = 0$) on the orientifold
side, so that $\frac{1}{g^2_a \mid_I} = \frac{1}{l_I}$ is indeed dual to
$\frac{1}{g^2_a \mid_H} = \frac{1}{l_H}$.}, the $D=10$ heterotic - type I
duality implies $l_H = l_I$, but this relation does no longer hold
at the one-loop level. The correct duality relation is found by expressing the
linear multiplets in terms of the chiral fields $S$ and $T_i$, and using the
duality dictionary derived by dimensional reduction of the $D=10$ 
duality relations \cite{Polchinski}
$\lambda_H = \lambda^{-1}_I$ and $M^2_H = \lambda^{-1}_I M^2_I$
(where $\lambda_H$, respectively $\lambda_I$, is the $D=10$ heterotic,
respectively type I dilaton). One finds $V_H = V_I$, where
$V_{H(I)} = \int d x^6 \scriptstyle{\sqrt{g^{(6)}_{H(I)}}}$ is the compact
volume in the string metric, as well as\footnote{Recall that
$(\mbox{Re} S)_H = V_H M^6_H / \lambda^2_H$ and
$(\mbox{Re} T_i)_H = V^i_H M^2_H$ (where $V^i$ denotes the volume of the
compact 2-torus $T^i$) on the heterotic side, and
$(\mbox{Re} S)_I = V_I M^6_I / \lambda_I$ and
$(\mbox{Re} T_i)_I = V^i_I M^2_I / \lambda_I$ on the orientifold side
\cite{Ibanez_orientifolds}.} $(\mbox{Re} S)_H = (\mbox{Re} S)_I$ and
$(\mbox{Re} T_i)_H = (\mbox{Re} T_i)_I$.

On the heterotic side, the expression of $l_H$ in terms of $S$ and $T_i$ is
given by the linear-chiral relation (\ref{eq:L_S_duality_loop}), which in the
particular case of the orbifold considered can be rewritten as
$\frac{1}{l} = \mbox{Re} S - \frac{b^H_{SU(12)}}{16 \pi^2}\,
\ln (V^{1/3}_H M^2_H)$, thus yielding:
\begin{equation}
  \frac{1}{g^2_a (\mu^2) \mid_H}\ =\ \mbox{Re} S\
  -\ \frac{b^H_{SU(12)}}{16 \pi^2}\: \ln\, (V^{1/3}_H M^2_H)\
  -\ \frac{b^H_a}{16 \pi^2}\: \ln \frac{\mu^2}{M^2_H}\ .
\label{eq:g_H_Z_ABD_2}
\end{equation}
This is however not yet the relevant expression for the low-energy gauge
couplings, since it holds in the trivial vacuum with a nonzero anomalous
$U(1)$ $D$-term, $D_X = \xi^2_H$. Considering the physical flat direction
with maximal gauge symmetry $G = SU(12) \times SO(8)$, along which $\xi^2_H$
is compensated for by vevs of the twisted moduli $M_k$, one finally finds:
\begin{eqnarray}
  \frac{1}{g^2_a (\mu^2) \mid_H}\ =\ \mbox{Re} S\ -\ \frac{b^I_a}{16 \pi^2}\:
  \ln\, (V^{1/3}_H \mu^2)\ .
\label{eq:g_H_Z_ABD_3}
\end{eqnarray}
The change from eq. (\ref{eq:g_H_Z_ABD_2}) to eq. (\ref{eq:g_H_Z_ABD_3}) is due
to the fact that along the flat direction considered, some twisted charged
states become massive and decouple, yielding
$b^H_{SU(12)} \rightarrow b^I_a$ and $b^H_a \rightarrow b^I_a$.
This results in a shift of the unification scale from the string scale to the
compactification scale $M_c = V^{1/3}_H$.

On the orientifold side, using the dilaton linear-chiral duality relation
(\ref{eq:l_S_duality}), we obtain:
\begin{equation}
  \frac{1}{g^2_a (\mu^2) \mid_I}\ =\ \mbox{Re} S\ +\ s_a\, m\
  -\ \frac{b^I_a}{16 \pi^2}\: \ln \frac{\mu^2}{M^2_I}\ .
\label{eq:g_I_Z_ABD_2}
\end{equation}
Unlike in the heterotic case, the Fayet-Iliopoulos term depends now on
the twisted moduli
\cite{Ibanez_anomalous,Poppitz,lln} and the point of maximal gauge symmetry
corresponds to $\xi^2_I =0$; as can be seen from an explicit linear-chiral
duality transformation, $\xi^2_I$ is proportional to the twisted modulus $m$
in the linear basis. Therefore, at the point of maximal symmetry, one has
$m=0$ and:
\begin{equation}
  \frac{1}{g^2_a (\mu^2) \mid_I}\ =\ \mbox{Re} S\
  -\ \frac{b^I_a}{16 \pi^2}\: \ln \frac{\mu^2}{M^2_I}\ ,
\label{eq:g_I_Z_ABD_3}
\end{equation}
which is obviously not dual to eq. (\ref{eq:g_H_Z_ABD_3}). Note that,
as was noticed in \cite{ABD},
duality would be restored if some (presumably nonperturbative)
mechanism ensured that $\xi^2_I =0$ for $\mbox{Re} M = 0$.


\section{More terms in the one-loop heterotic Lagrangian}

\subsection{One-loop corrections to the K\"ahler potential}

We now consider specific one-loop corrections to the K\"ahler potential
that are related to the 
well known holomorphic threshold corrections. In the remaining part of this 
paper we call these one-loop terms $\kappa$--corrections.
Both types of corrections are 
uniquely correlated  through the anomaly cancellation ten-dimensional 
Green-Schwarz terms \cite{GS}.

Let us start with  the perturbative heterotic string. 
As pointed out by Green and Schwarz for the gauge groups $E_8 \times E_8$ 
and $SO(32)$ the anomaly twelve form $I_{12}$ factorizes 
as $I_4 X_8$ and all gauge, gravitational
 and mixed gauge-gravitational 
anomalies are 
cancelled by the variation of well defined local counterterms. 
The crucial role in this cancellation is played by the 
antisymmetric field $B_{MN}$, which must transform nontrivially under gauge 
transformations. To recover classical gauge invariance at tree level the 
strength of the antisymmetric tensor field is modified so that it is gauge 
invariant and obeys $dH = I_4$. 
In the Horava-Witten model \cite{wh} representing the low-energy 
effective theory of the strongly coupled heterotic $E_8 \times E_8$ string, 
where the two $E_8$ gauge sectors 
live on opposite boundaries of the eleven-diemnsional bulk, 
the role of the effective antisymmetric tensor 
fields participating in the Green-Schwarz mechanism  is played 
by the components $C_{AB11}$ of the three-form $C$. 
 In general, there can be as many factorizable 
components of the anomaly form $I_{D+2}$ in $D$ dimensions, as many
antisymmetric tensor fields 
are available in the model under considarations. 
A generalized 
Green-Schwarz mechanism involving more than one antisymmetric tensor field 
is at work in Type I/Type IIB orientifold models \cite{Sagnotti_Z_3}, 
\cite{Ibanez_anomalous}. The local Green-Schwarz counterterms must be added to
the ten-dimensional action of the heterotic superstring, and must be 
taken into account 
when one performs the dimensional reduction/compactification down to 
four dimensions. Partial reduction of the GS terms has been performed in 
\cite{hpn1}, \cite{hpn2}, \cite{itoyama}, and more recently in 
\cite{bd}, \cite{lpt} in the context of the strongly coupled heterotic string. 
In particular reference \cite{itoyama} contains essentially the  
complete result. Among the terms coming from this reduction the best known
ones are 
the axionic parts of the holomorphic threshold corrections, $\pm \epsilon \, 
\theta F \tilde{F}$, where $\theta=Im(T)$,
which have exactly the same form in the strongly coupled 
and in the weakly coupled heterotic $E_8 \times E_8$ string cases (they 
can be directly computed as the large $T$ limit of the threshold corrections 
in weakly coupled orbifolds \cite{hpnss}, \cite{ss}). 
However, those are not the only relevant terms that contribute one-loop 
terms to the four dimensional Lagrangian. The other ones, ignored so far, 
are terms involving derivatives of the matter fields. To see these terms 
arising from the compactification we need the explicit form of the 
GS counterterms in the case of the 
$E_8 \times E_8$ heterotic string. If we denote the anomaly associated 
with the 
twelve-form $I_{12}$ by $G$, the new part in the 10d action satisfying 
$\delta_{gauge} S_{GS} + G = 0 $ can be written as 
\beq
S_{GS} = \int \left ( 4 ( \omega_{3L} - \frac{1}{30} \omega_{3Y} ) X_7 
- 6 B X_8 \right )
\label{eq5}
\eeq
where $d X_7 = X_8$ and $X_8 = \frac{1}{24} Tr F^4 - \frac{1}{7200} (Tr F^2 )^2
- \frac{1}{240} Tr F^2 tr R^2 + \frac{1}{8} tr R^4 + \frac{1}{32} (tr R^2)^2$.
To be specific, let us take the case of the standard embedding where 
$F^{(1)}_{MN}=R_{MN}$ and 
the Bianchi identity is fullfilled pointwise. The index on $F$ indicates that 
we use the gauge connection of only one of the two $E_8$'s, say the first one,
to solve the Bianchi identity.
As is well known, in that case the first $E_8$ is broken down to its $E_6$
subgroup, and the components of its connection with 
compact indices give rise to scalars $A$  in $h_{1,1}$ ${\bf 27}$ 
and $h_{1,2}$ ${\bf \bar{27}}$ representations of $E_6$. There are no matter 
fields associated with the unbroken $E_8$. The numbers $h_{a,b}$ are the Hodge
numbers of the Calabi-Yau manifold which forms the compact 6d space. 
The usual axionic thresholds come from the terms which couple compact 
components of B, the $B_{MN}$, to the $F^{(i)} \tilde F^{(i)}$ term 
composed of 4d gauge field strengths. These come from the integral of 
$-6 \, B \, X_8$ over the Calabi-Yau space. Noting that the expansion of the 
compact components of $B$ in harmonics on the CY space is 
$B_{MN} = \sum_{1}^{h_{1,1}} \theta^Z 
(x) \Omega^{Z}_{MN} (y)$ where $\Omega^Z $ are the harmonic $(1,1)$ forms,
the resulting 4d coupling is 
\beq
L_\theta = \frac{1}{20} \theta^Z \left ( F^{(1)} \tilde{F}^{(1)} 
\int_K \Omega^Z \wedge tr ( F^{(1)} \wedge F^{(1)}) \, - \, 
F^{(2)} \tilde{F}^{(2)} 
\int_K \Omega^Z \wedge tr ( F^{(2)} \wedge F^{(2)}) \right )
\label{eq6}
\eeq
i.e. the couplings have exactly the same magnitude and opposite sign
for $E_6$ and $E_8$ factors. The second coupling between 4d zero modes comes 
from the terms in $S_{GS}$ which contain space-time components of $B$ i.e. 
those proportional to $B_{\mu \nu}$. The physical degree of freedom associated
with these components is the pseudoscalar dual to $H=d B + \omega_{3L} 
- \frac{1}{30} \omega_{3Y}$. After integration by parts one obtains the
relevant part of the GS terms 
\beq
S_{GS, \, H} = - 6 \int_K  H X^{1}_{7}
\label{sgs}
\eeq
where the standard embedding is assumed and $X^{1}_{7} = \frac{1}{120} 
\omega^{1}_{3Y} ( tr F^{2}_{1} - \frac{1}{2} tr R^2 ) $. Let us note that if 
we were looking at corresponding terms with $\omega^{(2)}_{3Y}$ 
instead of $\omega^{(1)}_{3Y}$ then we would obtain the same expression with 
the opposite sign. The couplings of interest come from the terms 
\beq
\epsilon^{\mu \nu \rho M \sigma N PQRS} H_{\mu \nu \rho} Tr ({\cal A}_M 
\partial_\sigma 
{\cal A}_N) ( tr F^{2}_{1} - \frac{1}{2} tr R^2 )_{PQRS}
\label{haa}
\eeq
Assuming the Calabi-Yau space with $h_{1,2} = 0$, going over to complex 
coordinates ${\cal B}_1 = 1/ \sqrt{2} ({\cal A}_4 + i {\cal A}_5),..., 
{\cal B}_3=1 / \sqrt{2} 
({\cal A}_8 + i {\cal A}_9)$ and using the expansion \cite{redw} 
${\cal B}_1 =  T_{ax} A^{Kx} g^{a \bar{n} } \Omega^{K}_{1 \bar{n}}, \hfill
\bar{{\cal B}}_{\bar{1}} = \bar{T}_{\bar{a} \bar{x}} \bar{A}^{K\bar{x}} g^{\bar{a} n } \bar{\Omega}^{K}_{\bar{1} n}, \, etc.$ with 
orthogonality relation for the gauge group generators $Tr ( T_{a x} \bar{T}_{ \bar{b} \bar{y}} ) = g_{a \bar{b} } \delta_{x \bar{y}}$ one readily obtains 
the 4d coupling of the form 
\beq
\label{eqmix}
\delta L = \frac{i}{40} \epsilon^{\mu \nu \rho \sigma} H_{\mu \nu \rho}
( A^{Z x} \stackrel{\leftrightarrow}{\partial_\sigma} \bar{A}^{Y \bar{y}}
) \delta_{x \bar{y}} \int_K \sqrt{g_{(6)}} \epsilon^{MNPQRS}
g^{TU} \Omega^{Z}_{MT} \Omega^{Y}_{NU} ( tr F^{2}_{1} - \frac{1}{2} tr R^2 )_{PQRS} 
\eeq

In the next step one needs to use the duality relation between $H$ and the 
universal axion $D$: $H_{\mu \nu \rho} = g^{-1/2}_{4} e^{-6 \sigma} \phi^{3/2} 
\epsilon_{\mu \nu \rho \delta} \partial^{\delta} D $ where $g^{(10)}_{AB} = 
( e^{-3 \sigma} g^{(4)}_{\mu \nu}; \, e^\sigma g^{(0)}_{MN} )$ and $\phi $ is
related to the string dilaton\footnote{We recall that definitions of the
universal moduli complex scalars in the weakly coupled string convention are:
$S=e^{3 \sigma} \phi^{-3/4} + 3 i \sqrt{2} D$ and $T = e^{\sigma} \phi^{3/4} +
i \sqrt{2} \theta$.}. Then one obtains the terms $\partial^\mu D A 
\stackrel{\leftrightarrow}{\partial_\mu} \bar{A}$ 
which give rise to kinetic mixing of the dilaton superfield $S$ and untwisted
matter superfields 
$A$. To find the relation between the integrals which are coefficients in the 
terms (\ref{eqmix}) and (\ref{eq6}) we go now to the case $h_{1,1}=1$. 
Then the cohomology group $H^{1,1}$ consists only of the K\"ahler class 
generated by the K\"ahler form $k= i g_{i \bar{j}} dz^i \wedge d \bar{z}^{\bar{j}}$ where $g_{i \bar{j}}$ is the CY metric. Going over to the holomorphic 
coordinates one can see that the integrand in (\ref{eqmix}) contains the factor
\beq
g^{m \bar{n}} k_{\bar{q} m} k_{p \bar{n}} = - g_{p \bar{q}} = i k_{p \bar{q}}
\eeq
which finally allows one to write the coefficient of the 4d operator in 
(\ref{eqmix}) in the form of the 
topological integral $- \frac{1}{2} \int_K k \wedge tr (F^{(1)} 
\wedge F^{(1)})$ which is the same as the integral in the axionic threshold 
formula (\ref{eq6}). The simple derivation we have given here is equivalent 
to Witten's truncation on the six-torus when one replaces 
$g_{m \bar{n}} \rightarrow \delta_{m \bar{n}}$ and $k_{m \bar{n}} 
\rightarrow i \delta_{m \bar{n}} = \epsilon_{m \bar{n}}$. One should notice
that the antisymmetric symbol $\epsilon$ introduced above reduces to the usual
2d
antisymmetric symbol for each two-plane of the torus in the real coordinates. 

The above field-theoretical model calculation establishes the intimate
relation between the axionic threshold correction (which is imaginary part of
the holomorphic threshold correction) and the nonholomorphic correction
(\ref{eqmix}): either both of them vanish or both are present in the 4d
effective Lagrangian. In terms of the four dimensional 
chiral superfields the nonholomorphic correction leads to the modified 
K\"ahler potential 
\beq
K_S= - \log ( S + \bar{S} - \kappa A \bar{A}) 
\eeq
This mixing with the 4d dilaton superfield is expected to hold for all
untwisted matter multiplets, including the case of $h_{1,2} \neq 0$, 
in orbifold compactifications and is also valid in the case of  
nonstandard embeddings
of the gauge group. The corrections to the K\"ahler potential arise 
obviously at  one-loop order in the string coupling, since they come from the 
ten-dimensional Green-Schwarz terms. 
The sign of the mixing terms depends on the gauge group under consideration: 
it would be different for matter coming from the breaking of the second $E_8$. 
The same effect is present in the context of the 
Horava-Witten model which describes low-energy behaviour of the strongly 
coupled heterotic $E_8 \times E_8$ superstring. 
Let us complete this argument 
by summarizing the relevant calculation in the Horava-Witten model.
In that case the decomposition of the eleven-dimensional metric which leads 
to canonical Einstein-Hilbert terms in the four-dimensional action and 
exhibits relevant four-dimensional degrees of freedom is 
$g^{(11)}_{AB} = ( e^{- 2 \beta (x) - \gamma (x)} g^{(4)} ; e^{- 2 \beta (x) 
+ 2 \gamma (x)}; e^{\beta (x)} g^{(0)}_{IJ} )$. In this expression
$e^{ 3 \beta}$ is the fluctuating volume $V$ of the 6d Calabi-Yau space in
units of the reference volume $V_0 = \int_K \sqrt{g^{(0)}} $. 
The real parts of the chiral moduli superfields in the effective action 
are $Re(S) = e^{3 \beta}$ and $Re(T) = e^\gamma $. 
One should note, that the 10d source of kinetic terms for the 4d gauge fields
and for 4d scalars is the operator 
$ \frac{1}{ 8 \pi (4 \pi k_{11}^{2})^{2/3}} \int d^{10} x \sqrt{g}
Tr F_{AB} F^{AB}$. The relevant parts of it 
are 
\beq
 \frac{1}{ 8 \pi (4 \pi k_{11}^{2})^{2/3}}   \int d^{10} x \sqrt{g} 
Tr F_{\mu \nu } F^{\mu \nu} + 
  \frac{1}{ 4 \pi (4 \pi k_{11}^{2})^{2/3}}  \int d^{10} x \sqrt{g} 
Tr F_{\mu M} F^{\mu M} 
\label{ttm}
\eeq
where $\mu, \nu$ are noncompact and $M$ the compact indices, and $k_{11}^{2}$ 
is the 
eleven-dimensional gravitational constant. 
Using the decomposition of the gauge fields ${\cal A}_M$ given below formula
(\ref{sgs}) and 
the decomposition of the 11d metric in terms of $\beta, \, \gamma$
one obtains 
\beq
L_{kin} =  \frac{V_0}{ 8 \pi (4 \pi k_{11}^{2})^{2/3}}   
\int d^{4} x \sqrt{g^{(4)}} e^{ 3 \beta } Tr F_{\mu \nu } F^{\mu \nu} + 
\frac{V_0}{ 2 \pi (4 \pi k_{11}^{2})^{2/3}}   
\int d^{4} x \sqrt{g^{(4)}} \frac{6 }{e^{ \gamma }} |\partial_\mu A|^2 \, .
\eeq
As pointed out in \cite{wh} computing corrections to the effective 
Lagrangian in the linear order in perturbations of the metric of the 
compact six-dimensional space
corresponds to substitution 
$V = e^{3 \beta} V_0 \, \rightarrow 
V_{v,h} =  ( e^{3 \beta} \pm \xi_0 e^\gamma ) V_0$ at the visible and hidden 
walls respectively. The parameter $\xi_0$ is given by the topological integral
\beqa
\label{e:xpl} \xi_0 &=& - \frac{ \pi \rho_0 }{2 (4 \pi)^{4/3}
 k_{11}^{2/3}} \frac{1}{8 \pi^2} \int_X  k \, \wedge \, (\,
trF^{(1)} \wedge F^{(1)}
 - \frac{1}{2} \, trR \wedge R \,).
\eeqa 
The normalization that gives direct correspondence with the weakly coupled 
case is established through $V_0 = 2 \pi (4 \pi k_{11}^{2})^{2/3}$ and
$\alpha' = \frac{1}{(4 \pi )^{2/3} \pi^2 } \frac{k_{11}^{2/3}}{\rho_0} $
where one typically 
puts $\alpha' = 1/2$. 
In the kinetic terms of the gauge fields 
this leads to the well known difference between gauge couplings (gauge kinetic
functions) on different walls, and upon substitution into the second term in 
(\ref{ttm}) this gives immediately the corrections to the kinetic terms for
scalars\footnote{We remove the factor of $12$ through  rescaling of matter
fields $A$.}
\beq
L_{kin} = \frac{|\partial_\mu A |^2}{T + \bar{T}} \pm \xi_0 \frac{|\partial_\mu A_{v,h} |^2}{S + \bar{S}} .
\eeq
These corrections have opposite sign  
for matter on opposite walls, and are  
interpreted as corrections to the $S$-dependent part of the 4d K\"ahler 
potential
\beq
K_S = - \log ( S + \bar{S}) \pm \xi_0 \frac{A_{v,h} \bar{A}_{v,h}}{S + \bar{S}}
\approx -\log (S + \bar{S} \mp \xi_0 A_{v,h} \bar{A}_{v,h} ) 
\eeq
which is exactly the same expression that we have obtained in the 
weakly coupled heterotic string for  matter living on the visible wall 
(one should observe that with the standard 
normalization of the integrals $\xi_0 = \kappa$).  
As we can see clearly in the 
context of the Horava-Witten model these corrections are indeed uniquely 
related to the corrections to the holomorphic gauge kinetic functions 
$f_{v,h} = S \pm \xi_0 T$. They both have here the geometric interpretation 
of the correction to the volume of the 6d compact space induced by different 
vacuum gauge fluxes on different walls. In the weakly coupled regime the 
same effects are seen as one-loop quantum effects, and from the point of view 
of the four-dimensional effective theory the matching is exact. 
The reason for this matching 
is anomaly cancellation. The crucial observation is that the massless 
spectrum of weakly and strongly coupled theories is the same, and, hence, 
that ten-dimensional and eleven-dimensional Green-Schwarz 
terms participating in anomaly cancellation must be strictly related 
to each other. 
The most relevant term in the compactification down to four 
dimensions is the topological term $C \wedge G \wedge G$  of  eleven 
dimensional Horava-Witten Lagrangian, where $C$ is the eleven-dimensional 
three-form field and $G$ is its modified strength. It turns out 
that compactification\footnote{The compactification includes here 
substitution of the lowest order nontrivial solutions for $G$ along the 
eleventh dimension.} of the components with the index structure 
$\epsilon^{\mu \nu \rho \delta \, IJKL \, 11 MN} G_{\mu \nu \rho \delta}
G_{IJKL} C_{11 MN} $ produces in four dimensions 
axionic parts of the threshold corrections 
\beqa \label{e:23}
&\delta  L^{(4)} = 
 \frac{1}{ k_{11}^{2}} \frac{ \rho_0}{24 \pi} \left (
\frac{k_{11}}{4 \pi} \right )^{4/3}  tr(F^{(1)} \tilde{F}^{(1)})
\left [ \theta^{Z}  \int_X \Omega^{Z} \wedge ( tr(F^{(1)} 
\wedge F^{(1)})
- \frac{1}{2} tr(F^{(2)} \wedge F^{(2)}) \right . & \nonumber \\
& -  \left . \frac{1}{4} tr(R
  \wedge R ) ) \right ] &
\nonumber \\
&+ \frac{1}{ k_{11}^{2}} \frac{ \rho_0}{24 \pi} \left (
\frac{k_{11}}{4 \pi} \right )^{4/3} tr(F^{(2)} \tilde{F}^{(2)})
\left [ \theta^{Z}  \int_X \Omega^{Z} \wedge ( tr(F^{(2)} \wedge F^{(2)})
- \frac{1}{2} tr(F^{(1)} \wedge F^{(1)}) \right . & \nonumber \\
 &- \left . \frac{1}{4} tr(R \wedge R )) \right ] .&
\eeqa
These expressions, after substitution of $\rho_0$,
coincide with the axionic threshold corrections given in (\ref{eq6}).
In the same way the results of the compactification of the part of the 
$ C \wedge G \wedge G$ with the index structure 
$\epsilon^{M \rho N \delta \, IJKL \, 11 \mu \nu } G_{M \rho N \delta}
G_{IJKL} C_{11 \mu \nu} $
coincide precisely with the $\partial^\mu D A  
\stackrel{\leftrightarrow}{\partial_\mu} \bar{A}$ terms given 
in (\ref{eqmix}) for matter in the visible sector (the sign would be the
opposite one if we had matter in the hidden $E_8$ sector). 
Zero modes of the $C_{11 MN}$ and $C_{11 \mu \nu}$ 
coincide with the axions which are zero modes of $B_{MN}$ and $B_{\mu \nu}$.
Thus we can see, that the relation between holomorphic 
threshold corrections and  the $\kappa$-corrections to the K\"ahler 
potential receives even stronger support when viewed from the perspective 
of the strongly coupled heterotic models. There both types of 4d terms 
come from the topological term of the eleven dimensional supergravity,
which is uniquely constrained by supersymmetry and by anomaly cancellation.  
Either both types of terms are present, or both should be absent in any 
heterotic model. 
With the present observation 
taken into account the 4d effective Lagrangians from the weakly and strongly
coupled theories look exactly equivalent at the - respectively - one loop 
and linear in CY deformation orders\footnote{This concerns here models 
without five-branes in the 5d bulk on the strongly coupled side.}. 

Equally important from the point of view of the rest of the present paper 
is the observation, that it is natural to expect the corrections to 
$K_S$ to have the same nature also in the heterotic $SO(32)$ models. 
The structure of the Green-Schwarz terms is in this case very much the same
as described above and to be specific one can think of the decomposition 
$SO(32) \rightarrow SU(3) \times U(1) \times SO(26)$ where $SU(3)$ is 
identified with the holonomy group of the Calabi-Yau space
in analogy with the calculation presented earlier in this section.

\subsection{Modular transformations in the presence of  
 $\kappa$-corrections to the K\"ahler function}

As argued in the previous section in the (2,2) compactifications 
with holomorphic threshold corrections to the gauge couplings we expect 
the presence of nonholomorphic corrections which have a natural interpretation
of one-loop contributions to the K\"ahler function. 
In what follows we restrict 
ourselves to the stndard embedding, with matter in the visible $E_6$ sector 
only. Then
in the most symmetric case,
with just the single T-modulus, the relevant parts of the K\"ahler 
function and the kinetic function are 
\beq
K = - \log ( S + \bar{S} + \frac{3 \delta_{GS}}{4 \pi^2} 
 \log (T + \bar{T}) - \kappa A \bar{A}) - 3 \log ( T + \bar{T}) + 
\frac{A \bar{A}}{ T + \bar{T} } 
\eeq
\beq
f=S - (\frac{6 \kappa}{\pi}+ \frac{3}{2 \pi^2 } ) \log \eta^2 (T) +
\sigma (T) 
\eeq
where $ \kappa$ is a computable numerical parameter and $\sigma (T)$ is the 
holomorhic part of universal one-loop threshold correction which is invariant
under the $SL(2,Z)$ T-duality transformations and approaches
$-\frac{1}{4 \pi} T$ as $T \rightarrow \infty$. 
With the transformations 
$S \rightarrow S + \frac{3 \delta_{GS}}{4 \pi^2} 
 \log (i c T + d), \; T \rightarrow \frac{a T -i b}{i c T + d} $
the model given by the above $K$ and $f$ is no longer invariant at one loop.
One can cancel the variation of the $\kappa$-term in $K_S$ with 
a combination of the two modified transformations{}\footnote{This form of
 modified $T$-duality
transformation is reminiscent of the 
situation in the (2,2) orbifold models when 
the moduli C, charged under the subgroup H from the decomposition $H \times 
E_6 \times E_8$ of the orbifold gauge group are to obtain an expectation 
value. 
Then the transformation of T is no longer an $SL(2,Z)$ transformation, but 
becomes extended in a holomorphic way by terms which can be represented 
in the form of the power-law expansion in the blowing-up moduli C:
$ T \rightarrow \frac{a T - i b}{i c T + d} + g_n (T) C^n, \; n \geq 3 $
 \cite{flt}. 
One should notice, 
that when the duality transformation gets modified at tree level due to 
blowing up of the orbifold, either the transformation of S must get 
further modified at one loop, or the T-dependent counterterm under the 
logarithm must change to keep the $K(S,\bar{S})$ invariant.}:
\begin{itemize}
\item $S \rightarrow S + \frac{3 \delta_{GS}}{4 \pi^2} 
 \log (i c T + d)  + \gamma_S \, \kappa A \bar{A} ( \frac{1}{(i c T + d)(- i c \bar{T} + d) } 
-1 )/2 $,
\item $T \rightarrow \tilde{\Gamma} T = 
\frac{ a T - i b}{i c T + d} - \gamma_T \, 
\frac{ 2 \pi^2 \kappa}{3 \delta_{GS}} \frac{ (i c T + d)(- i c \bar{T} + d)
 -1}{(i c T + d )^2 (- i c \bar{T} + d )^2}
(T + \bar{T}) A \bar{A} $.
\end{itemize}
where $\gamma_S + \gamma_T = 1 $.
Any combination of $\gamma$'s is troublesome. First, let us note 
that the new terms should not be 
counted in $P_C K $ which enters the nonlocal term representing the triangle
graphs with a K\"ahler or sigma-model connection, as that would be a higher 
order effect. Second, the new terms turn a chiral superfield into a 
general superfield. Let us look more closely at various possibilities. 
The new terms in the transformation of S do not spoil the anomaly
cancellation, but they vary at one-loop order the physical, 1PI,
gauge coupling constant,
which is not consistent with the symmetry. These terms in the transformation
of T are proportional 
to the ratio of two parameters $\kappa$ and $\delta_{GS}$. 
One can assume that this ratio is  small and treat these new terms 
as one-loop contributions. Then one should drop it in the T-dependent 
threshold corrections to gauge couplings,  
but the K\"ahler function $K_T = - \log ( T + \bar{T} )$ 
becomes non-covariant, and there is no suitable term to restore 
covariance. One should observe for instance, that although the $T + \bar{T}$ 
and $|\eta (T) |^{-4}$ both transform identically with respect to the 
tree-level modular transformations, their derivatives are completly different, 
hence the higher order terms in the expansion in powers of $|A|^2$ of their 
images under the full transformations cannot be  matched to restore the 
covariance. \\
To see this more explicitly lets us assume the K\"ahler potential 
for the superfield T in the form 
\beq
K_T = - \log (T + \bar{T} + \beta  |\eta (T)|^{-4} )
\eeq
The requirement that the covariance is restored at one-loop gives a solution 
for the coefficient $\beta$ (we put here $\gamma_T =1$ for convenience)
\beq
\beta = - \frac{ 1 }{4 \pi } \frac{|\eta ( \Gamma T ) |^4 }{Re( G_2 
( \Gamma T ))}
\eeq
where $\Gamma T = (a T - i b)/(i c T + d)$. 
This solution obviously does not make sense, as the coefficient $\beta $ 
turns out to be a function depending on the parameters of the 
modular transformation. Let us note, that the form of the above solution 
suggests that one could try another counterterm, 
\beq
K_T = - \log (T + \bar{T} + \frac{\beta}{|\hat{G}_2 (T)|})
\eeq
where $\hat{G}_2$ is the covariant version of the Eisenstein form 
$G_2 (T)$, transforming with modular  
weight 2. It is a straightforward calculation to find out that this form of
the counterterm does not work, however, as under $\tilde{\Gamma}$ 
\beqa
&\hat{G}_2 (T) \rightarrow \hat{G}_2 (\tilde{\Gamma}T) = \hat{G}_2 (\Gamma T)
 \left ( 1 + \alpha A \bar{A} \left ( \frac{5}{24} (icT + d)^2 \frac{G_4(T)}{\hat{G}_2 (T) } + \right . \right . & \nonumber \\
& \left . \left . - \frac{1}{24} \left ( (i c T + d)^2 \frac{G_{2}^{2}(T)}{\hat{G}_2 (T) } - \frac{4 \pi^2 c^2}{ \hat{G}_2 (T)} - 4 \pi i  c (i c T + d) \frac{G_2 (T)}{ \hat{G}_2 (T) } + \frac{4 \pi }{(T + \bar{T} )^2 } \frac{ ( - i c \bar{T} + d )^2 }{\hat{G}_2 (T) } \right ) \right ) \right ) & 
\eeqa
with $\alpha = -
\frac{ 2 \pi^2 \kappa}{3 \delta_{GS}} \frac{ (i c T + d)(- i c \bar{T} + d)
 -1}{(i c T + d )^2 (- i c \bar{T} + d )^2}
(T + \bar{T})$. Hence, once again, one obtains a coefficient $\beta$ that 
depends explicitly on the 
transformation parameters. \\
Finally, one could ask whether, when we take $\gamma_S \neq 0$, 
in the gauge kinetic function $f$ the new term due to transformation of 
$S$, $\delta_{A S}$,  would not cancel against the term $\delta_{A T}$ from
the 
transformation of $T$. The point is that $\delta_{A S} \propto \delta_{GS}$ 
and $\delta_{A T} \propto \kappa^2 / \delta_{GS}$, hence the terms are of 
different orders in loop expansion parameters. In other words, such 
cancellation would require certain conspiracy between $\kappa$ and 
$\delta_{GS}$ -- quantities which have very different microscopic origin and 
in principle do not need to be related. In addition, even a successful 
cancellation in $f$ would not solve the problem of the covariance of $K_T$.    

The problem which we have described in this section does not imply
neccesarily 
that target-space duality does not hold in the heterotic string models. 
On the contrary, there are very good reasons to believe that it is an 
exact symmetry of many heterotic compactifications and as such should be 
representable in the effective Lagrangian. We would rather 
argue that there should exist further nonperturbative contributions
to the K\"ahler function for moduli and matter fields. To illustrate  
these statements, 
let us recall that we have identified the 
$\kappa$--corrections to the kinetic terms in field theoretical limit, 
i.e. in the limiting domain  of large $Re(T)$. In this limit 
the gauge kinetic 
function is $ f = S \pm \kappa T$ (for $E_6$ and $E_8$ sectors) 
which is at odds with the usual form of 
$T$-duality. To establish  $T$-duality one needs to promote $T$ in the $f$ 
to $\log \eta^2 (T)$ plus $SL(2,Z)$-invariant universal terms. 
This is the necessary extension of $f$ to the case
of arbitrary, both large and small,  values of $Re (T)$. Similar 
nontrivial extension of $K_S$ is likely to be necessary to restore one-loop 
duality in the full effective Lagrangian, valid over the whole moduli space.
The possible form of the generalized K\"ahler function could be 
\beq
K(S,\bar{S}; A, \bar{A}) = - \log ( S + \bar{S}\, - \kappa  A \bar{A} \, 
\frac{ \log ( j(T) - 744) + \log ( j( \bar{T} ) -744 )}{ 2 \pi (T + \bar{T})})
\label{res}
\eeq
where $j(T)$ is the $SL(2,Z)$-invariant form. 
With the expression (\ref{res}) substituted into the K\"ahler 
function the Lagrangian becomes modular invariant at the one-loop level 
without the need to modify the standard T-duality transformation 
for moduli and matter fields. Also, in the large $T$ limit it reduces to 
the expressions which we have computed in the section $4.1 \;$.  
The expression (\ref{res}) is not the only one which fulfills these 
conditions. There exist other possible choices\footnote{For instance: 
\beqa
&K(S,\bar{S}; A, \bar{A}) = - \log ( S + \bar{S}\, - \kappa  A \bar{A} \, 
|\eta (T) |^4 \, |j(T) -744|^{1/6} )\, ,& \nonumber \\
&K(S,\bar{S}; A, \bar{A}) = - \log ( S + \bar{S}\, - \kappa  A \bar{A} \, 
 | \hat{G}_2 (T) |   ( \frac{1}{3} - \frac{4}{ \log ( j(T) - 744) + 
\log ( j( \bar{T} ) -744 )})^{-1} / \pi^2 )\, .& \nonumber
\eeqa
to name just two examples. Also, these are lowest order terms in the expansion
in $A \bar{A}$. The exact form has to be determined by string 
calculations.}, with a different behaviour at small $T$. 

However, it still makes sense to discuss the effective Lagrangians 
with the perturbative form of $K_S$ (including $\kappa$-terms) and to 
compare them to perturbative Lagrangians coming from other 
compactifications, having in mind that one restricts oneself to a part of 
moduli space where $Re(T) \, >> \, 1$ in natural units and compare 
terms that might violate $T$-duality. 

The kinetic mixing between the dilaton modulus $S$ and matter 
fields transforming like tensors under $SL(2,Z)$ 
leaves some questions concerning heterotic--type IIB 
orientifold duality. 
On one hand it spoils the one-loop duality invariance of the naive effective 
Lagrangian, thus making the situation more symmetric between the two 
classes of models - T-duality could be violated at one-loop level
in a region of moduli space in both of the models. However, the way 
it would actually be violated seems to be  completely different on both sides. 
Moreover, we recall, that we believe that in orbifold models 
$\kappa$--corrections
are due to the existence of $N=2$ subsectors. Hence, in this respect,
making  
relations with orbifold models is justified 
in the analysis of the orientifolds like 
the $Z_{6}'$ one (whose dual partner is not known). 
In addition, we do not expect similar  matter-dilaton  
kinetic mixing in type IIB orientifold models. The reason is that there 
exists  no one-loop
coupling between the untwisted antisymmetric tensor field, corresponding to 
the imaginary part of the dilaton, and gauge fields. If the 
intuition gained during the analysis of the anomaly cancelling counterterms 
in heterotic models is correct, the kinetic mixing terms would rather 
be partners of 
the  $B^{(k)} \wedge F^{(k)}$ couplings. 
Speculating further, by the same token 
we would expect modifications of the form $\frac{1}{4}
(M_k + \bar{M}_k + \kappa_k A_k \bar{A}_k)^2$ to the K\"ahler functions 
of twisted moduli dual to twisted antisymmetric tensor fields, 
which have not been
seen in explicit calculations.


\section{Conclusions}

Many of the properties of string theory can be understood through a 
study of symmetries. In view of possible phenomenological
applications it is important to incorporate such symmetries
(if possible) into the low-energy effective field theory actions.
We have seen that target-space duality is a very useful symmetry, able 
to constrain severely those effective actions. In the framework of
simple compactification schemes of the heterotic theories, target-space
duality can be incorporated successfully. This even goes beyond 
the classical level and constrains the one-loop effective action,
including a mechanism for cancellation of all target-space
anomalies. 

In the first part of the paper we recall the discussion of
the simplest cases, $Z_3$ and $Z_7$ heterotic orbifolds, in detail.
We then try to extend this picture to certain type IIB
orientifold models (again $Z_3$ and $Z_7$), that have received some 
attention recently. One of the results of this paper is the
observation that target-space duality
now suffers from anomalies that cannot
be cancelled by a  mechanism similar to that
 in the heterotic case.
The failure at the level of sigma-gravitational anomalies could in principle 
be repaired by $T$-dependent corrections to the CP-odd $R^2$ terms,
but those would have to arise nonperturbatively.
Our results (independently of the question of interpretation) strengthen
the earlier arguments against a conjectured duality of a
certain class of heterotic-type IIB orientifold models. Target-space
dualities can be cancelled on the heterotic side, while no such
satisfactory 
field theoretical cancellation mechanism seems to be at work in 
the considered $Z_3$ and $Z_7$
type IIB orientifolds. 

Another face of the problem with T-duality, and consequently with
heterotic--type IIB orientifold duality, shows up when one tries to
interpret the threshold corrections that have been recently computed
in $Z_N$ orientifold models in the light of the 
cancellation mechanism suggested by the field-theoretical
analysis of anomalies.
In heterotic orbifolds, the structure of threshold corrections is
intimately linked to the mechanism of mixed sigma-model-gauge anomaly
cancellation, and one can actually infer from the explicit form of the
one-loop corrections to gauge couplings how these anomalies are precisely
compensated for.
The interpretation of threshold corrections in orientifolds is less
obvious due to the fact that the upper scale of logarithmic running is
not modular invariant. 
Upon performing the linear-chiral duality transformation that relates the
string result to the one-loop gauge couplings
computed in the effective supergravity theory, one finds that this results
in the impossibility to ensure target-space duality at the one-loop level:
either sigma-gauge anomalies are cancelled, but the twisted moduli K\"ahler
potential is not invariant; or the shift of the twisted moduli leaves an
uncancelled anomaly. Of course, these results
may simply point out that some nonperturbative terms in the 
orientifold Lagrangian are still missing, and in any case further string 
calculations in these models are truly needed.  However, as far as one can 
trust the field theory approach, we have seen that target-space symmetries
are a powerful tool to study and test conjectured relations between
various string theories and the structure of their effective Lagrangians.

It is well established that on the heterotic side, a symmetry 
that corresponds to $T$-duality 
does exist at the level of string
theory. Therefore, a suitable description of this symmetry should
be possible at the level of the field theoretic low-energy
effective Lagrangian. As we have seen earlier, this works without problems
in the simplest models. In section 4 we report on an attempt to
generalize this to more general cases (e.g. to those models which
contain $N=2$ subsectors).
We point out that a manifestly symmetric incorporation of
T-duality becomes problematic due to the appearance of one-loop
corrections to the K\"ahler potential. Several possibilities to make 
these terms consistent with T-duality are proposed. Here the question
arises, whether string theory can be approximated by a single unique
low-energy effective action. After all, the traditional approach
treated this action as relevant for the large T limit. Maybe 
different low-energy effective actions might be necessary 
to describe other patches of
$T$ moduli space. Nevertheless, we demonstrate that such a unique
description 
might be possible for the heterotic models under consideration.

\vskip 1cm

\noindent{\large \bf Acknowledgments}

\vskip .3cm

\noindent
The authors would like to thank I. Antoniadis, C. Bachas, E. Dudas,
M. Klein and N. Obers for useful conversations.
This work has been supported by TMR programs
ERBFMRX--CT96--0045 and CT96--0090.
Z.L. is additionaly supported 
by the Polish Committee for Scientific Research grant 2 P03B 037 15
and by M. Curie-Sklodowska Foundation.

\vfill
\eject


\vspace{1.5cm}
\appendix
\noindent{\large \bf Appendix A: Sigma-model anomalies in heterotic
orbifolds}
\setcounter{equation}{0} 
\renewcommand{\theequation}{A.\arabic{equation}}
\vspace{0.3cm}

In this appendix, we recall some basic facts about sigma-model anomalies in
$D=4$, $N=1$ heterotic orbifolds. At tree level, the Lagrangian is invariant
under $SL(2,{R})$ reparametrizations of the geometric
moduli\footnote{We consider only the diagonal moduli $T_i$, $i=1,2,3$, which
are common to all orbifolds. In the presence of off-diagonal moduli
$T_{i \bar j}$ ($i \neq j$), the target-space duality group is actually larger
than $\prod_{i=1}^{3} SL(2,{R})_{T_i}$.} \cite{flat, lmn, flt},
\begin{equation}
  T_i\ \rightarrow\ \frac{a_i T_i - i b_i}{i c_i T_i + d_i}\ ,  \hskip 2cm
  a_i d_i - b_i c_i = 1\ ,
\label{eq:SL_2_R_T}
\end{equation}
together with linear transformations of the matter fields:
\begin{equation}
  \Phi_{\alpha}\ \rightarrow\ \prod_{i=1}^3\,
  (i c_i T_i + d_i)^{n^i_{\alpha}}\, \Phi_{\alpha}\ ,
\label{eq:SL_2_R_Phi}
\end{equation}
where the $n^i_{\alpha}$ are the modular weights of the chiral superfield
$\Phi_{\alpha}$.
At the one-loop level, this symmetry known as target-space duality
is generally anomalous; indeed, it acts as a chiral rotation on fermions
and can have mixed anomalies with gauge symmetries and
gravity\footnote{Strictly speaking, as explained in subsection 2.2,
target-space duality transformations combine a sigma-model reparametrization
and a K\"ahler transformation, both of which are anomalous
\cite{Derendinger_sigma}. Following the
literature, we shall call the resulting anomalies (\ref{eq:sigma_gauge}) and
(\ref{eq:sigma_grav}) indifferently ``target-space
duality anomalies'' or ``sigma-model anomalies''.}. Under an $SL(2,{R})_{T_i}$
transformation, the one-loop Lagrangian undergoes an anomalous variation
\begin{equation}
  \delta {\cal L}_{\rm anomaly}\ =\ \frac{\theta_i}{32 \pi^2}\
  \sum_a\, b^{\prime i}_a\, F^a \widetilde{F}^a\ -\ \frac{\theta_i}{768 \pi^2}\
  b^{\prime i}_{grav}\, R \widetilde{R}\ ,
\end{equation}
where $\theta_i = \arg (i c_i T_i + d_i)$ is the angle of the chiral rotation,
and the mixed sigma-gauge and sigma-gravitational anomaly coefficients
 $b^{\prime i}_a$ and $b^{\prime i}_{grav}$ are given by
\cite{Derendinger_sigma,lidl}:
\begin{eqnarray}
  b^{\prime i}_a & = & -\, C_2 (G_a)\, +\, \sum_{\alpha}\,
  (1 + 2 n^i_{\alpha})\, T(R_{\alpha})\ ,  \label{eq:sigma_gauge}  \\
  b^{\prime i}_{grav} & = & 21\ +\ 1\ +\ b^{\prime i}_{mod}\ -\
  \dim G\ +\ \sum_{\alpha}\, (1 + 2 n^i_{\alpha})\ .  \label{eq:sigma_grav}
\end{eqnarray}
In Eq. (\ref{eq:sigma_gauge}), $C_2 (G_a)$ is the quadratic Casimir of the
gauge group $G_a$ and $T(R_{\alpha})$ is the index of the representation
$R_{\alpha}$ of $G_a$; in Eq. (\ref{eq:sigma_grav}), $\dim G$ is the dimension
of the total gauge group, $21$ and $1$ stand for the contribution of the
gravitino and the dilatino respectively, and $b^{\prime i}_{mod}$ denotes the
contribution of the other (gauge singlet) modulinos, which is model-dependent.
The mixed sigma-gauge anomaly is reproduced by the variation of the following
nonlocal term \cite{cardoso1,Derendinger_sigma}:
\begin{equation}
  {\cal L}_{\rm n.l.}\ =\ -\, \frac{1}{32 \pi^2}\: \sum_a \int \!
  \mbox{d}^2 \theta\ W^a W^a\, \sum_i\, b^{\prime i}_a\ P_C \left[\,
  \ln (T_i + \bar T_i)\, \right]\ +\ \mbox{h.c.}\ ,
\label{eq:L_nl}
\end{equation}
where $P_C = -\, \frac{1}{16}\, {\Box}^{-1} \bar D^2 D^2$ is the chiral
projector, defined such that $\bar D\, (P_C H) = 0$ for any superfield $H$,
and $P_C H = H$ if $H$ is a chiral superfield. In terms of components,
${\cal L}_{\mbox{n.l.}}$ also contains a non-harmonic contribution to gauge
couplings:
\begin{equation}
  -\, \frac{1}{4}\, \sum_a\, F^a F^a\, \left(\, -\, \sum_i\,
  \frac{b^{\prime i}_a}{16 \pi^2}\: \ln (T_i + \bar T_i)\, \right)\ .
\end{equation}
This shows that non-harmonic one-loop corrections to the gauge couplings are
related, through supersymmetry, to the presence of sigma-model anomalies.

Since the discrete version of the above symmetries is related to $T$-duality,
which is an exact symmetry of the heterotic string, these anomalies must be
compensated in some way. Two mechanisms can be at work
\cite{Derendinger_sigma}: (i) a Green-Schwarz mechanism \cite{GS} similar to
the one responsible for the cancellation of abelian gauge anomalies \cite{dsw},
which involves a
non-linear transformation of the dilaton superfield at the one-loop level:
\begin{equation}
  S\ \rightarrow\ S\ -\ \frac{1}{8 \pi^2}\ \delta^i_{GS}
  \ln (i c_i T_i + d_i)\ ;
\label{eq:S_shift}
\end{equation}
(ii) non-invariant, $T_i$-dependent holomorphic corrections to the gauge
kinetic function. Such corrections arise from loops of massive string states
and are associated with complex planes that are left invariant by some of
the orbifold twists (corresponding to $N=2$ subsectors of the orbifold).
In a large class of models, they take the form \cite{DKL}
\begin{equation}
  \Delta f^{\rm 1-loop}_a\ =\ -\, \frac{1}{4 \pi^2}\: \sum_i\, c_{a, i}\,
  \ln \left[ \eta (T_i) \right]\ ,
  \label{eq:threshold}
\end{equation}
where the coefficient $c_{a, i}$, to be determined by an explicit string
computation, vanishes when the $i^{\rm th}$ complex plane is rotated by all
twists. Note that $T_i$-dependent threshold corrections explicitly break the
continuous $SL(2,{R})_{T_i}$ symmetry to its discrete version
$SL(2,{Z})_{T_i}$, while the Green-Schwarz mechanism preserves
it. Under an $SL(2,{Z})_{T_i}$ transformation, one has $\eta^2 (T_i) 
\rightarrow (i c_i T_i + d_i)\, \eta^2 (T_i)$ and
\begin{equation}
  f^{\rm 1-loop}_a\ \rightarrow\ f^{\rm 1-loop}_a\ -\ \frac{1}{8 \pi^2}\:
  \left(\, \delta^i_{GS} + c_{a, i}\, \right)\, \ln (i c_i T_i + d_i)\ .
\end{equation}
Anomaly cancellation occurs provided the following relations are satisfied:
\begin{equation}
  b^{\prime i}_a\ =\ \delta^i_{GS}\, +\, c_{a, i}\ .
\label{eq:b'_i_a}
\end{equation}
Since $c_{a, i} = 0$ when the $i^{\rm th}$ complex plane is rotated by all
twists, Eq. (\ref{eq:b'_i_a}) imposes a strong constraint on the corresponding
sigma-model anomalies, which must then be gauge-group independent (exactly as
happens for abelian gauge anomalies compensated by a Green-Schwarz mechanism).
Cancellation of mixed sigma-gravitational anomalies is realized in a similar
manner.

Collecting all contributions to the one-loop effective Lagrangian for the
gauge fields, Eq. (\ref{eq:L_nl}) and (\ref{eq:threshold}), and
using Eq. (\ref{eq:b'_i_a}), one obtains \cite{Derendinger_sigma}:
\begin{eqnarray}
  {\cal L}_{\rm gauge} & = & \frac{1}{4}\: \sum_a \int \! \mbox{d}^2 \theta\
  W^a W^a\ P_C \left\{\, \left[\, S + \bar S\ -\ \frac{1}{8 \pi^2}\
  \delta^i_{GS} \ln (T_i + \bar T_i)\, \right] \right.  \nonumber \\
  & &  \left. -\ \frac{1}{8 \pi^2}\ \sum_i\,
  (b^{\prime i}_a - \delta^i_{GS})\, \ln \left[\, |\eta (T_i)|^4
  (T_i + \bar T_i)\, \right]\, \right\}\ +\ \mbox{h.c.}\ .
\label{eq:L_GK_loop}
\end{eqnarray}
Eq. (\ref{eq:L_GK_loop}), together with Eq. (\ref{eq:K_S_loop}), shows that
the structure of one-loop corrections to the effective string action is
strongly constrained by target-space duality anomaly cancellation. Note that
anomaly considerations do not tell us anything about possible holomorphic,
modular invariant corrections \cite{hpnss} that are not included in
Eq. (\ref{eq:L_GK_loop}).


\vspace{1cm}
\appendix
\noindent{\large \bf Appendix B: Green-Schwarz mechanism in the linear
multiplet formalism}
\setcounter{equation}{0} 
\renewcommand{\theequation}{B.\arabic{equation}}
\vspace{0.3cm}

The Green-Schwarz mechanism that cancels part of the sigma-model anomalies
in heterotic orbifolds can be naturally described in the linear 
multiplet formalism\footnote{For a useful reference about the linear multiplet
in effective heterotic string theories, see \cite{Derendinger_linear} and
references therein.}.
Indeed, in terms of the string massless states, the
axion-dilaton-dilatino system fits into a linear multiplet
$L = (l, B_{\mu \nu}, \chi)$, where the antisymmetric two-tensor
$B_{\mu \nu}$ is dual to the model-independent axion $a = \mbox{Im} S$,
$\partial_\mu a \sim \epsilon_{\mu \nu \rho \sigma}\, \partial^{\nu}
B^{\rho \sigma}$. $L$ couples to the gauge fields in such a way that the
combination $\widehat L = L - 2 \Omega$, where $\Omega$ is the Chern-Simons
superfield defined by $\bar D^2 \Omega = \sum_a W^a W^a$ and
$D^2 \Omega = \sum_a \bar W^a \bar W^a$, is gauge invariant. A gauge-invariant
Lagrangian for $L$ then takes the simple form ${\cal L}_L = \int \!
\mbox{d}^4 \theta\ \Phi (\widehat L)$. The transformation to the dual 
formulation in terms of the
dilaton chiral superfield is accomplished by treating $\widehat L$ as an
unconstrained superfield and adding the constraint
$\bar D^2 (\widehat L + 2 \Omega) = D^2 (\widehat L + 2 \Omega) = 0$
to the Lagrangian:
\begin{equation}
  {\cal L}\ =\ \int \! \mbox{d}^4 \theta\ \left[\, \Phi (\widehat L)\
  -\ \frac{1}{2}\: (S + \bar S)\, (\widehat L + 2 \Omega)\, \right]\ ,
\label{eq:L_S_lagr}
\end{equation}
where $S$ is the dilaton chiral superfield, and an unconstrained  
superfield $\Sigma$
defined by $S = \bar D^2 \Sigma$ plays the role of the Lagrange multiplier.
The equation of motion for $\Sigma$ is nothing else but the constraint for
$\widehat L$, while the equation of motion for $\widetilde L$ gives the duality
relation:
\begin{equation}
  \frac{\partial \Phi}{\partial \widehat L}\ =\ \frac{1}{2}\: (S + \bar S)
  \hskip 1cm \Rightarrow \hskip 1cm \widehat L\ =\ \widehat L\, (S, \bar S)\ .
\label{eq:L_S_duality}
\end{equation}
Putting (\ref{eq:L_S_duality}) into (\ref{eq:L_S_lagr}), one obtains the
Lagrangian for $S$, ${\cal L}_S = \int \! \mbox{d}^4 \theta\: K (S, \bar S)
+ \frac{1}{4}\: \int \! \mbox{d}^2 \theta\: f(S)$, with $f(S) = S$ and the
K\"ahler potential given by:
\begin{equation}
  K\, (S, \bar S)\ =\ \left[\, \Phi (\widehat L)\ -\ \frac{1}{2}\:
  (S + \bar S)\, \widehat L\, \right]_{\ \widehat L\ =\ \widehat L\,
  (S, \bar S)}\ .
\end{equation}
At tree level, one has $\Phi (\widehat L) = \ln \widehat L$, which leads to the
duality relation $1 / \widehat L = (S + \bar S) / 2$ and the K\"ahler potential
$K (S, \bar S) = - \ln (S + \bar S)$. The Green-Schwarz terms needed for
the cancellation of sigma-model anomalies come at one loop and take
the form $\Delta {\cal L}_{GS} = \sum_i \frac{\delta^i_{GS}}{16 \pi^2}\:
\widehat L\, \ln (T_i + \bar T_i)$. In components, this contains a coupling
between the antisymmetric tensor $B_{\mu \nu}$ and the sigma-model connection.
Through a linear-chiral duality transformation, $\Delta {\cal L}_{GS}$
translates into a one-loop mixing between the dilaton and the geometric
moduli. Indeed, the one-loop duality relation reads
\begin{equation}
  \frac{1}{\widehat L}\ =\ \frac{1}{2}\ \left[\, S + \bar S\ -\
  \frac{1}{8 \pi^2}\ \sum_i\, \delta^i_{GS} \ln (T_i + \bar T_i)\, \right]\ ,
\label{eq:L_S_duality_loop}
\end{equation}
implying a K\"ahler potential of the form (the dots refer to further
corrections that are not related to sigma-model anomaly cancellation):
\begin{equation}
  K^{\rm 1-loop}\, (S, \bar S)\ =\ -\, \ln \left[\, S + \bar S\
  -\ \frac{1}{8 \pi^2}\ \sum_i\, \delta^i_{GS} \ln (T_i + \bar T_i)\
  +\ \ldots\, \right]\ .
\label{eq:K_S_loop}
\end{equation}
Target-space modular invariance then requires that $S$ transforms at one loop
according to (\ref{eq:S_shift}).

In type IIB orientifold compactifications we need to add more linear 
multiplets $m_k$, $k=1,...,n_f$ where $f$ is the number of twisted sectors. 
Such an extension of the linear multiplet formalism becomes 
somewhat subtle in the context of supergravity. To illustrate the trouble,
let us take the simple case of just a single additional linear multiplet. 
If one neglects the superpotential couplings, the Lagrangian is given by 
${\cal L}= S_0 \bar{S}_0 \Phi (\hat{L}, \hat{m})$ where $S_0 = (z_0, \psi_0, 
f_0)$ is the conformal compensator. Let us take 
\beq
\Phi = \frac{1}{\sqrt{2}} e^{K/3} ( - X^{-1/2} + \frac{s}{2} X^{1/2} Y^2 )\ ,
\label{eqphi}
\eeq
with $X= e^{K/3} \frac{\hat{L}}{S_0 \bar{S}_0 }, \; 
Y= e^{K/3} \frac{\hat{m}}{S_0 \bar{S}_0 } $. Using this form of $\Phi$ 
one obtains the graviton kinetic term in the form
\beq
- \frac{1}{2} R \, ( \frac{ |z_0|^3 }{(2 e^K l )^{1/2} } + \frac{s m^2}{4} 
\frac{(2 e^K l )^{1/2}}{|z_0|^3} )\ .
\eeq
It is clear that choosing the value of $z_0$ which corresponds to the 
Einstein frame, $|z_0|^3 = (1 - \frac{s}{4} m^2)(2 e^K l)^{1/2}$ for small 
$m$, leads in general to a mixing between $m$ and $l$.
The effective Lagrangian becomes simple and similar in its form to the 
heterotic effective Lagrangian only in the vicinity of the point $<m>=0$. 
There the choice (\ref{eqphi}) gives in the leading order in $m$ 
the gauge coupling $\frac{1}{g^2} = \frac{1}{l} + s \, m$ and the quadratic 
form of the K\"ahler potential for the dual chiral field $M$: 
$K \sim (M + \bar{M})^2 + ... \, $. In this regime it is also possible 
to take into account one-loop corrections to $1/ g^2$, e.g. through 
$\delta_{(1)} \Phi \sim \frac{3 b_0}{4} \frac{ \hat{m}}{ S_0 \bar{S}_0} 
\log (X)$ which produces $ \sim b_0 \log ( l) $ in $1/ g^2$.       

\vfill
\eject


\end{document}